\documentclass[11pt,a4paper,twoside]{article}
\usepackage[utf8]{inputenc}
\usepackage{cite}

\usepackage{amsmath}

\usepackage{amssymb}

\usepackage[english]{babel}

\usepackage{epsfig}

\usepackage{a4wide}

\usepackage{graphics, color}

\def\rhs{rhs\ }

\title{Quantum spin chains of Temperley-Lieb type:
  periodic boundary conditions, spectral multiplicities and finite temperature}

\author{Britta Aufgebauer \ and Andreas Kl\"umper \\
 \parbox{0.9 \linewidth}{\vspace{0.4 \baselineskip}\centering
  Fachbereich C -- Physik, Bergische Universit\"at Wuppertal, \\ 42097
    Wuppertal, Germany}}
\date{\today}

\begin{document}
\maketitle
\begin{abstract}
We determine the spectra of a class of quantum spin chains of Temperley-Lieb
type by utilizing the concept of Temperley-Lieb equivalence with the $S=1/2$
$XXZ$ chain as a reference system. We consider open boundary conditions and in
particular periodic boundary conditions. For both types of boundaries the
identification with $XXZ$ spectra is performed within isomorphic
representations of the underlying Temperley-Lieb algebra. For open boundaries
the spectra of these models differ from the spectrum of the associated $XXZ$
chain only in the multiplicities of the eigenvalues. The periodic case is
rather different. Here we show how the spectrum is obtained sector-wise from
the spectra of globally twisted $XXZ$ chains. As a spin-off, we obtain a
compact formula for the degeneracy of the momentum operator eigenvalues. Our
representation theoretical results allow for the study of the thermodynamics
by establishing a {\em TL-equivalence at finite temperature and finite field}.
\end{abstract}

\section{Introduction}
Since the introduction of the Temperley-Lieb algebra \cite{TemperleyLieb71},
the concept of Temperley-Lieb equivalence has been widely used in statistical
mechanics, see for instance \cite{BaxterBook82,PMBook91} and references
therein. The original motivation of this concept was the computation of
physical properties of the $Q$-states Potts model on the self-dual line by a
mapping to the six-vertex model. The possibility of such a mapping is
interesting as the configuration spaces of the Potts model and of
the six-vertex model are rather different. The underlying mechanism of this
mapping is of algebraic and representation theoretical type and allows for
relating the eigenvalues of the transfer matrix of the Potts-model to those of
the six-vertex model.  Needless to say, the concept of Temperley-Lieb
equivalence has also attracted strong attention in mathematics \cite{Jones83}.

By now, there are many more models like the $RSOS$-models \cite{ABF84}, the
graph-models \cite{OwczarekBaxter87,Pasquier87} and certain vertex-models
\cite{KlumperBiq89,Kulish03} which are based on representations of the
Temperley-Lieb algebra. These models allow for explicit evaluations of some of
their physical properties by a mapping to the six-vertex model. Also, these
models are integrable in the traditional sense, because the local interactions
satisfy the Yang-Baxter equation as a consequence of the Temperley-Lieb
relations.

The Temperley-Lieb equivalence naturally extends to the quantum counterparts
of the above statistical mechanical models, such as the quantum $RSOS$ models
and quantum spin-$S$ chains \cite{Parkinson88,BarberBatchelor89,KlumperBiq89},
all of which are related to the spin-1/2 Heisenberg chain with partial
anisotropy ($XXZ$ chain).

The concept of Temperley-Lieb equivalence has been established for systems
with open boundary conditions
\cite{TemperleyLieb71,BaxterBook82,PMBook91}. Many physical properties do not
depend on the boundary conditions if the thermodynamical limit is
taken. According to the Temperley-Lieb equivalence, transfer matrices or
Hamiltonians of models based on the Temperley-Lieb algebra have the same
spectrum (up to degeneracies of the eigenvalues) as the corresponding
operators of the `standard reference' six-vertex model or the $XXZ$ quantum
spin chain. Obviously, properties that depend on the special type of boundary
conditions, like finite size-data yielding conformal dimensions, or properties
that depend on multiplicities, like thermodynamics of the quantum chains, are
not covered!

For some cases, notably the critical $RSOS$-models and their quantum
counterparts, the entire spectrum is known, because the underlying Hilbert
space is lower-dimensional than in the case of the standard reference
model. The $RSOS$-models with periodic boundary conditions allow for an
analysis based on the fusion algebra \cite{BazhanovReshetikhin89}, which is
rather different from a representation theoretical treatment of for instance
the periodic Temperley-Lieb algebra \cite{PeriodicTL93,PeriodicTL94}.  The
Bethe ansatz like eigenvalue equations for the $RSOS$-models look like those
of the $XXZ$ chain with special twisted boundary conditions. The question
about degeneracies of eigenvalues is simply answered with one or zero.

In this paper we are interested in a general approach utilizing representation
theoretical concepts to tackle the outlined problems and we are going to apply
our approach to quantum chains with higher dimensional spins. Here the
question about degeneracies of eigenvalues finds rather different answers than
in the case of the $RSOS$ models. In fact, the eigenvalues are rather highly
degenerate. Actually, for ferromagnetic exchange interactions, the quantum
chains exhibit residual entropy, i.e.~the degeneracy of the ground-state
increases exponentially with the chain length. For periodic boundary
conditions, we find Bethe ansatz like equations with twist angle taking values
from a much larger set than in the case of the $RSOS$ models. Here the twist
angles comprise real and imaginary values! We believe that our results
complete the studies of the spectral problem of the so-called biquadratic
spin-1 chain and generalizations
\cite{Parkinson88,BarberBatchelor89,KlumperBiq89,KlumperBiq90a,KlumperBiq90b,AKS1992,KS1994,KS1996}. Despite
the large number of papers devoted to the spectral problem of this model, even
conscientious coordinate Bethe ansatz calculations did not yet reveal the high
degeneracies \cite{KS1994,KS1996}.

The outline of the article is as follows. In section \ref{TL-models} we
introduce the class of quantum spin chains we are going to address.  In
section \ref{open boundaries} the case of open boundaries is discussed. The
multiplicities of the eigenvalues are obtained using representation theory of
the Temperley-Lieb algebra.  The emphasis of this paper is on section
\ref{periodic boundaries} where we deal with periodic boundaries. Here the
determination of the multiplicities is more involved, because the spectrum is
sector-wise obtained from the spectra of several $XXZ$-chains with different
twisted boundary conditions. We use representations of the periodic
Temperley-Lieb algebra \cite{PeriodicTL93,PeriodicTL94} which are constructed
from translationally invariant reference states with zero and non-zero
momentum eigenvalues. For these models the physical properties of the
anti-ferromagnetic ground state and a few excited states have been reported in
the literature \cite{Parkinson88,KS1994,KS1996}. Here, we present the complete
treatment of the entire spectrum, in particular for the system with periodic
boundary conditions. Finally, in section \ref{applications} we discuss the
thermodynamical properties of the biquadratic spin-1 chain which turn out to
be rather different from those of the related $XXZ$ chain.

\section{Temperley-Lieb quantum spin chains: open and periodic boundaries}\label{TL-models}
The Temperley-Lieb algebra $TL_N(\lambda)$ is the unital associative algebra
over $\mathbb{C}$ generated by $e_1,\,e_2,\,\dots e_{N-1}$ with relations
(\ref{TL-relations}), depending on the complex parameter $\lambda$
\begin{equation}
\begin{split}
  e_i^2  = & \lambda \; e_i, \quad \, \mbox{for}\quad i=1,\,2,\dots,\,N-1,  \\ 
 e_i \; e_{i+1} \; e_i   = &   e_i, \quad \quad \mbox{for} \quad i=1,\,2 ,\dots ,\,N-2, \\
 e_i \; e_{i-1} \; e_i  = &   e_i, \quad \quad \mbox{for} \quad i=2,\,3,\dots ,\,N-1,  \\
 e_i\;e_j = & e_j\;e_i, \;\;\; \mbox{for}\quad |i-j|> 1. \label{TL-relations}
\end{split}
\end{equation} 
The periodic Temperley-Lieb algebra $PTL_N(\lambda)$ has one more generator
 $e_N$ additional to the generators of $TL_N(\lambda)$, and in addition
 to (\ref{TL-relations}) also the relations (\ref{per TL-relations}) hold
\begin{equation}
\begin{split}
   e_N^2 = & \lambda \; e_N, \\ 
 e_N \; e_i \; e_N   =&    e_N, \quad \quad \mbox{for} \; i=1, N-1,\\ 
 e_i \; e_N \; e_i   = &   e_i, \,\;\quad \quad \mbox{for} \; i=1,N-1,
\\  
 e_N\;e_i  = & e_i\;e_N, \;\:\:\, \mbox{for}\;i\neq 1,N-1.
\end{split}\label{per TL-relations}
  \end{equation}
In contrast to the algebra $TL_N(\lambda)$, which is finite
dimensional, the algebra $PTL_N(\lambda)$ is infinite
dimensional for $N>2$\cite{Levy91}.

\subsection{Spin chains of Temperley-Lieb type with open boundary conditions}
The global Hilbert space $\mathcal{H}_N$ of an $N$-site spin-$S$ chain is
typically given as the $N$-fold tensor product
\begin{equation*}
\mathcal{H}_N=h_1 \otimes h_2 \otimes \cdots \otimes
h_N\quad\mbox{with}\quad h_i = \mathbb{C}^{2S+1}\quad \mbox{for}\;
i=1,\dots, N.
\end{equation*} 
For a given representation $\rho$ of the algebra $TL_N(\lambda)$ on
$\mathcal{H}_N$
\begin{equation}
\begin{split}
\rho: TL_N(\lambda)&\longrightarrow \mbox{End}(\mathcal{H}_N)\\
e_i&\mapsto b_i,\label{TLrep}
\end{split}
\end{equation}
the Hamiltonian of the associated $N$-site TL spin chain with open boundaries
is given by
\begin{equation}
H^o=\sum\limits_{i=1}^{N-1}b_i. \label{open Hamiltonian}
\end{equation}
For the construction of $TL_N(\lambda)$-representations on $\mathcal{H}_N$ we
consider the algebra $U_q(sl_2)$, generated by $S^+, S^-$ and $ q^{\pm S^z}$
under the relations \cite{KulishReshetikhin81}
\begin{equation}
q^{S^z}S^{\pm}q^{S^z}=q^{\pm 1}S^{\pm},\quad \left[S^+,S^-\right ]=\frac{q^{2S^z}-q^{-2S^z}}{q-q^{-1}},
\end{equation}
with $S^x, S^y, S^z$ the spin operators and $S^\pm=S^x \pm i S^y$.  The local
Hilbert space $\mathbb{C}^{2S+1}$ is a $(2S+1)$-dimensional highest-weight
representation of $U_q(sl_2)$. Let
\begin{equation}
B_S = \left \{\left |M \right >\,:\,M=-S,-S+1,\dots,S \right \}\label{spinbasis}
\end{equation}
be the basis of $\mathbb{C}^{2S+1}$ with 
\begin{eqnarray}
S^\pm\;\left |M \right >\;&=&\;\sqrt{(S\pm M+1)(S\mp M)}\;\left |M\pm 1 \right >, \\
S^z\;\left |M \right >\;&=&\;M\;\left |M \right >.
\end{eqnarray} 
$\mathcal{H}_N$ is a $U_q(sl_2)$ representation via iterated use of the
coproduct $\Delta$
\begin{equation}
\Delta(q^{\pm S^z})=q^{\pm S^z}\otimes q^{\pm S^z},\quad
\Delta(S^\pm)=q^{ S^z}\otimes S^\pm +S^\pm \otimes q^{-S^z}.
\end{equation}
We obtain a representation (\ref{TLrep}) of $TL_N(\lambda)$ for 
\begin{equation}
b_i = id^{\otimes i-1}\otimes P \otimes id^{N-(i+1)}\quad\mbox{with}\quad
P=\left|\Psi\right>\left<\Psi\right|\in\mbox{End}(h\otimes h)\label{bi}
\end{equation}
being the projector onto the two-site $U_q(sl_2)$ spin-zero singlet
\begin{equation}
\left|\Psi\right
>=\sum_{{M_1,M_2=-S}\atop{M_1+M_2=0}}^{S}(-1)^{S-M_1}q^{-M_1}\left|M_1\right>\otimes\left|M_2\right>.\label{Uqsinglet}
\end{equation}
According to the well known realisation of $TL_N(\lambda)$ as a diagram
algebra we introduce the following graphical notation for the operators $b_i$:
\begin{equation}\mbox{\parbox{0.5cm}{$ b_i  $}
\hfill $=$ \hfill\parbox{7.5cm}{$\underbrace{~\epsfig{file=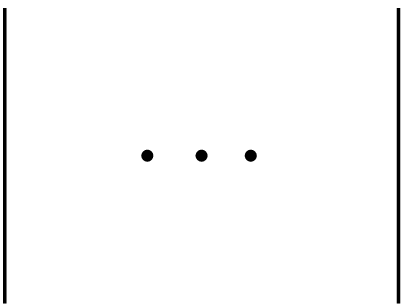,
      width= 2 cm}~}_{id^{\otimes
      (i-1)}}~\quad\epsfig{file=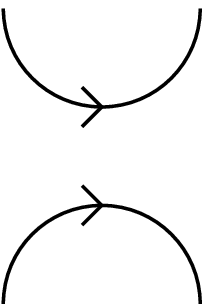, width= 1 cm}~ \;\quad
  \underbrace{~\epsfig{file=2nid.eps, width= 2 cm}~}_{id^{\otimes
      (N-i-1)}} $} }\label{bgra}
\end{equation}
The vector $\left|\Psi\right>$ and its dual $\left<\Psi\right|$ are depicted
as
\begin{equation}
\mbox{
\parbox{1cm}{$\left |\Psi\right >$}
\hfill ${=\quad}$ \hfill
\parbox{3 cm}{ $~\epsfig{file=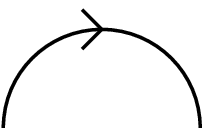, width= 1 cm}~$}\hfill and $\quad\quad$\hfill\parbox{1cm}{$\left<\Psi\right|$}
\hfill ${=\quad}$ \hfill
\parbox{3 cm}{ $~\epsfig{file=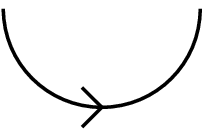, width= 1 cm}~$}}
\end{equation}
With the usual Hermitian scalar product on $\mathcal{H}_N$ and $q=r e^{i\phi}$
($r,\phi\in\mathbb{R}$) we find
\begin{equation}
 \lambda=\left< \Psi \,\right |\left. \Psi \right >=\sum\limits_{i=-S}^{S}r^{2i}\ge 2S+1.\label{genericTL}
\end{equation}
For these values of the TL-parameter the algebra $TL_N(\lambda)$ is
semisimple (the so-called generic case).  In order to allow for arbitrary
values of $\lambda$, in particular for non-generic TL-parameters, $q$ has to
be considered as a formal variable with respect to the bilinear form on
$\mathcal{H}_N$ (see \cite{PasquierSaleur89}).  We concentrate our discussion
on the generic case (\ref{genericTL}) and comment on the non-generic case,
leading to critical spin-chains, in sections \ref{nongenericO} and
\ref{nongenericP}.  For $q=1$ the local projection operator in terms of
spin-operators is given by
\begin{equation}
\left|\Psi\right>\left<\Psi\right| =\left <\Psi \right.\left |\Psi \right > \prod_{J=1}^{2s} \left
  [1-\frac{\left ( \vec{S}_1 + \vec{S}_2 \right )^2}{J(J+1)} \right ] \label{Projektor}
\end{equation} 
For arbitrary $q$ it takes the form
\begin{equation}
\left|\Psi\right>\left<\Psi\right| =\left <\Psi \right.\left |\Psi \right > \prod_{J=1}^{2s} \left
  [1-\frac{\Delta(C)}{\left
  [J+1/2\right ]^2_q -\left [1/2 \right ]^2_q} \right ], \label{Projektor2}
\end{equation} 
with $C$ the Casimir-Operator 
\begin{equation}
C=S^-S^++\left (
  \frac{q^{S^z+1/2}-q^{-(S^z+1/2)}}{q-q^{-1}}\right)^2-\left( \frac{q^{1/2}-q^{-1/2}}{q-q^{-1}}\right)^2
\end{equation}
and $\left [x\right ]_q=(q^x-q^{-x})/(q-q^{-1}) $.  Note that the physically
most interesting Hamiltonians differ from (\ref{open Hamiltonian}) by a
negative scale factor. 
\newline 
For dimension $d:=2S+1=2$ of the local Hilbert space, in the generic case the
representation of $TL_N(q+q^{-1})$ on $\mathcal{H}_N$ generates
$\mbox{End}_{U_q(sl_2)}\mathcal{H}_N$ and vice versa (double centralizer
property). For $d>2$ the algebra which is the centralizer of the
$TL_N(\lambda)$-representation (\ref{bi}) and vice versa has been constructed in
\cite{KulishManojlovicNagy08}.

\subsection{Periodic boundaries}
We obtain a representation of $PTL_N(\lambda)$ on $\mathcal{H}_N$ by mapping
the first $N-1$ generators according to (\ref{TLrep}) and the additional
generator $e_N$ to $b_N$ acting on $h_N \otimes h_1$ as the local projection
operator $P$ and elsewhere as identity.  The TL-Hamiltonian for periodic
boundaries takes the form
\begin{equation}
H^p=\sum\limits_{i=1}^{N}b_i. \label{Hamiltoniansp}
\end{equation}
Let the map $\alpha: h\rightarrow h$ be defined as
\begin{equation}
\mbox{\parbox{5.5cm}{$ \alpha \quad= \quad\left (\mbox{id}\otimes \left <\Psi\right | \right )\circ\left (\left |\Psi\right >\otimes\mbox{id}\right )$}
\hfill $=\quad$ \hfill\parbox{4cm}{
$~\epsfig{file=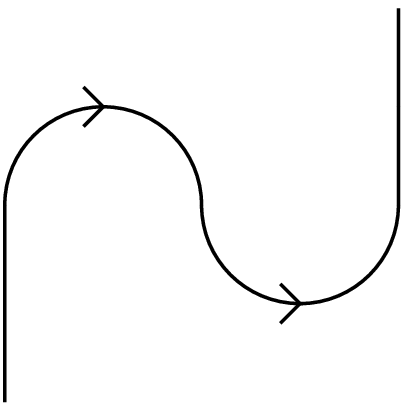, width= 2 cm}~$}}\label{Alpha}
\end{equation}
With respect to the basis (\ref{spinbasis}) we find for (\ref{Uqsinglet})
\begin{equation}
\alpha\; :\;\left|M\right>\mapsto (-1)^{d-1}e^{-2i\phi
  M}\left|M\right> \label{alphaformula}. 
\end{equation}
For $\phi\neq 0$ the Hamiltonian (\ref{Hamiltoniansp}) realizes globally
twisted periodic boundary conditions (with total twist angle $N\phi$). 

For $k\in \mathbb{N}$ we define
\begin{equation}
 I:=\prod_{i=1}^{k}b_{2i}\quad\mbox{and}\quad J:=\prod_{i=1}^{k}b_{2i-1}.
\end{equation}
For $N=2k$ we find for our representations the additional relations
\begin{equation}
 I\,J\,I=\left[\mbox{tr}(\alpha^{N/2})\right ]^2\,I\quad \mbox{and}\quad J\,I\,J=\left[\mbox{tr}(\alpha^{N/2})\right ]^2\,J
\end{equation}
meaning that for even $N$ we are dealing with representations of a finite
dimensional quotient of $PTL_N(\lambda)$ (compare \cite{PeriodicTL93} and
\cite{Nichols06}).  For $N=2k+1$ we find
\begin{equation}
I\,J\,b_N\,I=\left[\alpha^{N}\otimes \mbox{id}^{\otimes (N-1)}\right ]\,I\quad \mbox{and}\quad
J\,b_N\,I\,J\,b_N=\left[\mbox{id}^{\otimes (N-1)}\otimes\alpha^{N}\right ]\,J\,b_N.\label{oddN}
\end{equation}

\subsection{The $S=1/2$ $XXZ$ reference model \label{XXZ}} 
The TL-operators $b_i$ for the $XXZ$ chain are obtained from (\ref{Uqsinglet})
for $S=1/2$. The 2-site projector for $ q\in\mathbb{R}$ is given by
\begin{equation}
\left|\Psi\right >\left<\Psi\right|=\Bigl(q^{-1/2}\left|+-\right>-q^{1/2}\left|-+\right>\Bigr)\Bigl(q^{-1/2}\left<+-\right|-q^{1/2}\left<-+\right|\Bigr).\qquad
\label{XXZprojektor}
\end{equation}
The interaction of the $i$th with the $(i+1)$th spin is described by the
local Hamiltonian
\begin{equation}
h_{i,i+1}=\left(\frac{q+q^{-1}}{4}-b_i\right).
\end{equation}
\subsubsection{Open boundaries}
The $N$-site $XXZ$ Hamiltonian for open boundaries is given by 
\begin{equation}
\begin{split}
H^o_{XXZ}=\sum_{i=1}^{N-1}h_{i,i+1}=&\sum_{i=1}^{N-1}\Bigl(S_i^+S_{i+1}^-+S_i^-S_{i+1}^++2\Delta S_i^zS_{i+1}^z\Bigr)\\&+\frac{1}{2}(q-q^{-1})(S_1^z-S_{N}^z).\\
\end{split}
\label{XXZopen}
\end{equation}
The anisotropy parameter $\Delta$ of the
$XXZ$ chain is related to $\lambda$ via 
\begin{equation}
\lambda=2\Delta=q+q^{-1}.\nonumber
\end{equation}
The spectrum of the Hamiltonian (\ref{XXZopen}) is known by the Bethe ansatz
within eigenspaces of the magnetization operator 
$S^z_{\rm tot}:=\sum_{i=1}^{N}S^z_i$.
Apart from a trivial shift, the Hamiltonian (\ref{open Hamiltonian}) is given
by the same algebraic expression in terms of $TL(\lambda)$ generators as the
Hamiltonian (\ref{XXZopen}). Thus the spectrum of (\ref{open Hamiltonian}) is
equal to the spectrum of the $XXZ$ Hamiltonian with $\Delta=\lambda/2$ within
equivalent $TL_N(\lambda)$-subrepresentations. For $\lambda$ in the semisimple
regime the global Hilbert space $\mathcal{H}_N$ decomposes into a direct sum
of irreducible $TL_N(\lambda)$-representations. It is then convenient to use
these irreducibles to identify the spectra. Each type of irreducible
$TL_N(\lambda)$-representation occurs in the Hilbert space of the $XXZ$ chain,
because the corresponding $TL_N(\lambda)$-representation is faithful.

\subsubsection{Periodic boundaries}
The $XXZ$ Hamiltonian for periodic boundaries is given by
\begin{equation}
H^p_{XXZ}=\sum_{i=1}^{N-1}h_{i,i+1}+h_{N,N+1}=\sum_{i=1}^{N}\Bigl(S_i^+S_{i+1}^-+S_i^-S_{i+1}^++2\Delta S_i^zS_{i+1}^z\Bigr)
\label{XXZperiodic}
\end{equation}
as $h_{N,N+1}=(q+q^{-1})/{4}-b_N$ with
\begin{equation}
S_{N+1}^{\pm}:=S_1^{\pm}\quad \mbox{and}\quad S_{N+1}^{z}:=S_1^{z}. \label{periodicb}
\end{equation}
Globally twisted boundaries for twist angle $\phi$ can be obtained by changing
only the 2-site projector of the operator $b_N$ to
\begin{equation}
 \Bigl(e^{-i\phi/2}q^{-1/2}\left|+-\right>-e^{i\phi/2}q^{1/2}\left|-+\right>\Bigr)\Bigl(e^{i\phi/2}q^{-1/2}\left<+-\right|-e^{-i\phi/2}q^{1/2}\left<-+\right|\Bigr).
\end{equation}
In this case the boundary conditions for (\ref{XXZperiodic}) are given by
 \begin{equation}
S_{N+1}^{\pm}:=e^{\pm i\phi}S_1^{\pm}\quad \mbox{and}\quad S_{N+1}^{z}:=S_1^{z}. \label{twisti}
\end{equation}
Alternatively one may introduce an angle $\phi/N$ for each $b_i$. The resulting Hamiltonian is equivalent to the above one by a
simple similarity transformation. 
The twist angle $\phi$ enters the Bethe ansatz equations (given here for
$\Delta > 1$)
\begin{equation}
e^{i\phi}\left(\frac{\sinh(v_l+i\frac{\eta}{2})}{\sinh(v_l-i\frac{\eta}{2})}\right)^N=\prod_{{j=1}\atop{ j\neq l}}^{k}\frac{\sinh(v_l-v_j+i\eta)}{\sinh(v_l-v_j-i\eta)},\quad\quad \Delta=\cosh(\eta)\label{BAnsatz1}
\end{equation} 
for the Bethe ansatz rapidities $v_l$ with $1\le l\le k$ in the $XXZ$ sector
with $s^z=N/2-k$.  The eigenvalue of the Hamiltonian (\ref{XXZperiodic}) for a
solution $(v_1,\dots,v_k)$ of (\ref{BAnsatz1}) is given by
\begin{equation}
 E=\frac{N}{2}\cosh(\eta)+\sum\limits_{j=1}^k\frac{\sinh^2(\eta)}{\sinh(v_j+i\frac{\eta}{2})\sinh(v_j-i\frac{\eta}{2})}\label{BAnsatz2}.
\end{equation}
To obtain the spectra of the Hamiltonians (\ref{Hamiltoniansp}) we construct
$PTL(\lambda)$-subrepresentations equivalent to $S^z_{\rm tot}$ eigenspaces of
the $XXZ$ Hamiltonian with twisted boundaries.

\section{Invariant subspaces for open boundary conditions}\label{open boundaries}
We show how the irreducible $TL_N(\lambda)$-representations are constructed in
the global Hilbert space $\mathcal{H}_N$ of a given TL-model and determine
their multiplicities. The results follow directly from the representation
theory of the algebra $TL_N(\lambda)$. They are important for the analysis of the periodic case in section
\ref{periodic boundaries}. \newline
The formula for the multiplicities (\ref{opendim}) was obtained earlier in \cite{Kulish03} and \cite{KulishManojlovicNagy08} using representation theory of the centralizer algebra of the $TL_N(\lambda)$-representation on $\mathcal{H}_N$ .  
\subsection{Representation theory of $TL_N(\lambda)$}\label{RepTL}
We briefly summarize the essentials of the representation theory of the
algebra $TL_N(\lambda)$ necessary for our treatment. We keep our account
short, for more details we refer the reader to the references
\cite{PMBook91,Goodman1989,Westbury93}. 
The algebra
$TL_N(\lambda)$ is semisimple iff
\begin{equation}
P_k(\lambda)\neq 0 \quad \mbox{for}\quad 1\le k < N. 
\end{equation}
The polynomials $P_k$ are defined recursively via (\ref{polynomials})
\begin{equation}
\begin{split}
P_0(x) & =  1 \,,\quad P_1(x)  =  x,\\
P_k(x) &= x\; P_{k-1}(x)-P_{k-2}(x), \quad \mbox{for}\quad k\ge 2.
\label{polynomials} 
\end{split}
\end{equation}
The zeros of $P_k$ are real with absolute value not larger than 2. They are given by 
\begin{equation}\label{zeros}
x_l = 2\, \mbox{cos}\left(\frac{l \pi}{k+1}\right)\quad
\mbox{for}\quad l=1, 2, \dots, k.
\end{equation}
For $\lambda$ in the semisimple regime the isoclasses of irreducible
representations of $TL_N(\lambda)$ are parameterized by $k \in
\mathbb{N}$, $0\le k\le \left[N/2 \right]$. The square brackets denote
the largest integer equal to or smaller than the argument. The
$TL_N(\lambda)$-representation corresponding to $k$ will be denoted by
$O(N,k)$. The dimensions are given by 
\begin{equation}
\mbox{dim}\left(O(N,k)\right)=
\begin{cases}
1,\quad &\mathrm{for}\quad k=0,\\
\frac{1}{k+1}{ N\choose k},\quad &\mathrm{for}\quad k=N/2,\quad
(N\, \mbox{even})\\
{ N\choose k}-{N\choose k-1},\quad &\mathrm{else}. 
\end{cases}\nonumber
\end{equation}
\begin{figure}
\centering
\includegraphics[width=10cm]{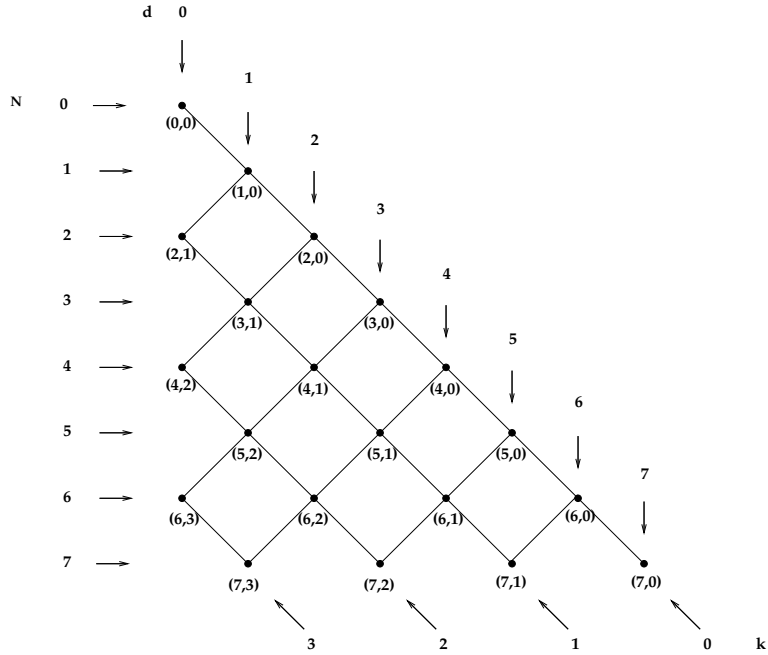}
\caption{Bratelli diagram of $TL(\lambda)$}
\label{Brat}
\end{figure}
An important tool for our analysis is the decomposition rule for
irreducibles of $TL_N(\lambda)$ into irreducibles of the subalgebra
$TL_{N-1}(\lambda)$ which is:

\begin{equation}
O(N,k)\downarrow_{TL_{N-1}}\cong
\begin{cases}
O(N-1,0),\quad &\mathrm{for}\quad k=0,\\
O(N-1,k-1),\quad &\mathrm{for}\quad k=N/2,\\
O(N-1,k)\oplus O(N-1,k-1), \quad &\mathrm{else}.
\end{cases}\label{dec}
\end{equation}
This decomposition rule may be read off from the Bratelli diagram, see
figure~\ref{Brat}.

\subsection{Construction of $TL_N(\lambda)$-representations in the generic case}\label{opensectors}
In order to construct the irreducible representations of $TL_N(\lambda)$ in
the global Hilbert space $\mathcal{H}_N$ of the $N$-site chain we define the
space
\begin{equation}
\Omega_N:=\left\{ \omega_N\in\mathcal{H}_N\;:\;b_i\;\omega_N =\;0\;\;\mbox{for}\;1 \le i\le N-1\right\}.
\end{equation}
The dimension of $\Omega_N$ is the multiplicity of the one-dimensional trivial
representation of $TL_N(\lambda)$ in $\mathcal{H}_N$. Below we will prove that
the multiplicity of the representation $O(N,k)$ for $0\le k\le \left[N/2
  \right]$ in $\mathcal{H}_N$ is equal to the dimension of the space
$\Omega_{N-2k}$.
 
\begin{figure}
\centering
\includegraphics[width=12cm]{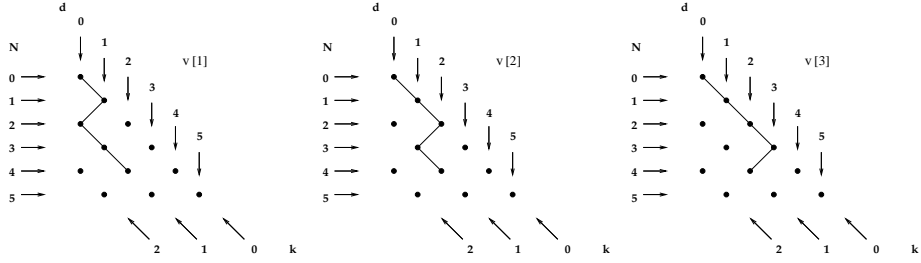}
\caption{The generating paths for $O(4,1)$}\label{pathpicture}
\end{figure}

A representation of type $O(N,1)$ in $\mathcal{H}_N$ is constructed starting
from the vector
\begin{equation}
\mbox{b}[1; \omega_{N-2}]:=\Psi\otimes\,\omega_{N-2},\label{Start1}
\end{equation}
with $\omega_{N-2} \in \Omega_{N-2}$ arbitrary. Acting with the TL-operators
on this initial state one finds that the vectors
\begin{equation}
\mbox{b}[i; \omega_{N-2}]:= b_{i}\,
b_{i-1}\cdots b_2\, \mbox{b}[1; \omega_{N-2}],\quad 1\le i \le N-1,
\end{equation}
span a $TL_N(\lambda)$-invariant subspace. An orthogonal basis for this
subspace is given by
\begin{equation}
\begin{split}
\mbox{v}[1; \omega_{N-2}]:&=\mbox{b}[1;\omega_{N-2}],\\
 \mbox{v}[i; \omega_{N-2}]:&=\frac{P_{i-1}(\lambda)}{P_{i}(\lambda)}
 \left [b_i\,\mbox{v}[i-1; \omega_{N-2}]-\frac{P_{i-2}(\lambda)}{P_{i-1}(\lambda)}\mbox{v}[i-1; \omega_{N-2}]\right]\quad \mbox{for}\quad 1 < 
 i \le N-1. \label{vbasis}
\end{split}
\end{equation}
The operators $b_i$ act as
\begin{equation}
\begin{split}
b_1\,\mbox{v}[1; \omega_{N-2}]&=P_1(\lambda)\,\mbox{v}[1; \omega_{N-2}],\\
\left .\begin{array}{r}b_i\,\mbox{v}[i-1; \omega_{N-2}]\\b_i\,\mbox{v}[i; \omega_{N-2}] \end{array}\right \}&=\frac{P_{i-2}(\lambda)}{P_{i-1}(\lambda)}
 \,\mbox{v}[i-1; \omega_{N-2}]+\frac{P_{i}(\lambda)}{P_{i-1}(\lambda)}\mbox{v}[i; \omega_{N-2}]=\frac{P_1(\lambda)}{P_{i-1}(\lambda)}\mbox{b}[i,\omega_{N-2}],\\
b_i\,\mbox{v}[j; \omega_{N-2}]&=0\quad\mbox{else}.
\end{split}\label{b action}
\end{equation}
In particular the vector v$[N-1; \omega_{N-2}]$ yields a
$TL_{N-1}(\lambda)$-representation of type $O(N-1,0)$. By induction over $N$
it follows that the constructed representation is indeed irreducible and that
$\left\{ \mbox{b}[i; \omega_{N-2}]:\; 1\le i \le N-1 \right\}$ and $\left\{
\mbox{v}[i; \omega_{N-2}]:\; 1\le i\le N-1 \right\}$ are bases.
\newline
Generalizing the construction to arbitrary $k$, a representation of type 
$O(N,k)$ is constructed starting from the vector 
\begin{equation}
\mbox{b}[1,3,5,\cdots, 2k-1; \omega_{N-2k}]:=\Psi^{\otimes k}\otimes\,\omega_{N-2k},\label{Start}
\end{equation}
with $\omega_{N-2k}\in\Omega_{N-2k}$. The ${N\choose k}-{N\choose k-1}$ many
vectors
\begin{equation}
\mbox{b}[i_1,i_2,\cdots, i_k; \omega_{N-2k}]:=\lambda^{-k}\left[\prod_{l=1}^{i_1}b_{l}\right]\left[\prod_{l=3}^{i_2}b_{l}\right]\cdots
      \left[\prod_{l=2k-1}^{i_k}b_{l}\right]\mbox{b}[1,3,5,\cdots, 2k-1; \omega_{N-2k}],
\label{basis k}
\end{equation}  
with indices subject to
\begin{equation}
2k-1\le i_k\le N-1 \quad\mbox{and}\quad 2l-1\le i_l< i_{l+1}\quad
\mbox{for all}\quad l<k  \label{basis ka}
\end{equation}
span a $TL_N(\lambda)$-invariant subspace. The order of the products in
(\ref{basis k}) is such that the indices increase from right to left.
Orthogonal basis vectors with the restriction (\ref{basis ka}) on the indices
$i_1, i_2, \dots, i_k$ are constructed recursively via
\begin{equation}
\begin{split}
&\mbox{v}[i_1,\cdots,i_l,\cdots, i_k;
\omega_{N-2k}]\\&:=\frac{P_{{i_l-1}}(\lambda)}{P_{{i_l}}(\lambda)}\Bigl (
b_{i_l}\,\mbox{v}[i_1,\cdots,i_l-1,\cdots,i_k; \omega_{N-2k}]
-\frac{P_{{i_l}-2}(\lambda)}{P_{{i_l-1}}(\lambda)}\,\mbox{v}[i_1,\cdots,i_l-1,\cdots,i_k;
\omega_{N-2k}]\Bigr ). \label{vasis k}
\end{split}
\end{equation} 
The $TL_{N-1}(\lambda)$-representation $O(N-1,k-1)$ is spanned by the vectors
(\ref{vasis k}) with $i_k=N-1$.  The vectors (\ref{vasis k}) can be identified
with the set of decreasing paths on the Bratelli-diagram connecting the points
$(0,0)$ and $(N,k)$. Associated with the vector v$[i_1,i_2,i_3,\cdots, i_k;
  \omega_{N-2k}]$ is the decreasing path through the points $(0,0)$,
$(i_1,0)$, $(i_1,1)$, $(i_2,1)$, $(i_2,2)$ $\cdots$ $(i_k,k)$, $(N,k)$. Let
$(i, k)$ be a path-point of the vector v. Let this point have the vertical
index $d$. The action of $b_i$ on the vector v is determined by the location
of the two path-points with horizontal indices $i-1$ and $i+1$. If these two
points have different vertical indices the vector belongs to the kernel of
$b_i$. If the two points have vertical index $d-1$ we find
\begin{equation}
\begin{split}
 b_i\; \mbox{v}&=\frac{P_{d-2}(\lambda)}{P_{d-1}(\lambda)}\;\mbox{w} + \frac{P_d(\lambda)}{P_{d-1}(\lambda)}\;\mbox{v},\quad \mbox{for}\;d\neq 1,\\
b_i\; \mbox{v}&=\lambda\;\mbox{v},\quad \mbox{for}\;d = 1,
\end{split}
\end{equation}
where w is the vector belonging to the path obtained by replacing the
point $(i, k)$ of the path of v by the
point $(i, k+1)$. If the $(i-1)$th and the $(i+1)$th point have both the
vertical index $d+1$ the operator $b_i$ acts as
\begin{equation}
 b_i\;\mbox{v}=\frac{P_d(\lambda)}{P_{d+1}(\lambda)}\;\mbox{v}+\frac{P_{d+2}(\lambda)}{P_{d+1}(\lambda)}\;\mbox{w}
\end{equation}
with w being obtained by replacing the point $(i, k)$ by the point $(i, k-1)$.
The paths for the $TL_4(\lambda)$-representation $O(4,1)$ are given as an
example in figure~\ref{pathpicture}.

\subsection{Dimension of $\Omega_{N}$}\label{Dimoff}
With the initial conditions $\Omega_0:=\mathbb{C}$ and $\Omega_1:=h$,
$d:=dim(h)=2S+1$ we find
\begin{equation}
\mbox{dim}(\Omega_N)=P_{N}(d),\label{opendim}
\end{equation}
by induction: from the TL-relations we find the inclusion
\begin{equation}
\Omega_N \subset \Omega_{N-1}\otimes h\nonumber
\end{equation}
for $\mathcal{H}_N$ - subspaces. The space $\Omega_N$ is the kernel of the map
\begin{equation}
b_{N-1}:\Omega_{N-1}\otimes h\longrightarrow \Omega_{N-2}\otimes \left|\Psi\right> \label{Abbildung}
\end{equation}
For $\omega_{N-2}\in\Omega_{N-2}$ consider the representation $O(N,1)$
constructed from $\omega_{N-2}$. From equation (\ref{b action}) it follows
\begin{equation} \mbox{v}[N-1; \omega_{N-2}]\in \Omega_{N-1}\otimes h\nonumber
\end{equation} 
and also
\begin{equation}
b_{N-1}\,\mbox{v}[N-1;
\omega_{N-2}]=\frac{P_1(\lambda)}{P_{N-2}(\lambda)}\mathrm{b}[N-1; \omega_{N-2}]\neq 0.
\end{equation}
This proves the surjectivity of the map (\ref{Abbildung}) and we obtain the
recursive dimension formula:
\begin{equation}
\mbox{dim}(\Omega_N)=d\,\mbox{dim}(\Omega_{N-1})-\mbox{dim}(\Omega_{N-2})\label{open dimension}
\end{equation}
which coincides with (\ref{polynomials}). An explicit formula for $d > 2$ is
given by
\begin{equation}
\mbox{dim}(\Omega_N)=\frac{\left ( d + \sqrt{d^2-4} \right )^{N+1}- \left ( d -
 \sqrt{d^2-4} \right )^{N+1}}{2^{N+1}\sqrt{d^2-4}}.
\end{equation}
This formula shows that the dimension of the space $\Omega_N$ grows
exponentially with $N$ for $d>2$.  For the $XXZ$ representation ($d = 2$) we
have
\begin{equation}
\mbox{dim}(\Omega_N)=N+1. 
\end{equation}
Each eigenspace of the operator 
$S^z_{\rm tot}$ in the $XXZ$ representation is a direct sum of irreducible
$TL_N(\lambda)$-representations as follows:
\begin{equation}
\mbox{Eig}(s^z=\pm(N/2-k), \mathcal{H}_N)\cong O(N,k)\oplus O(N,k-1)\oplus
\cdots \oplus O(N,0).\label{sz eigenraum}
\end{equation}
Decomposing the global Hilbert space $\mathcal{H}_N$ of a given TL-model into
a direct sum of $XXZ$-$S^z_{\rm tot}$ eigenspaces the multiplicity of (\ref{sz
  eigenraum}) in $\mathcal{H}_N$ is equal to
\begin{equation}
 \mbox{dim}(\Omega_{N-2k})-\mbox{dim}(\Omega_{N-2k-2}).\label{sz multi}
\end{equation}

\subsection{The non-generic case\label{nongenericO}}
\begin{figure}
\centering
\includegraphics[width=10cm]{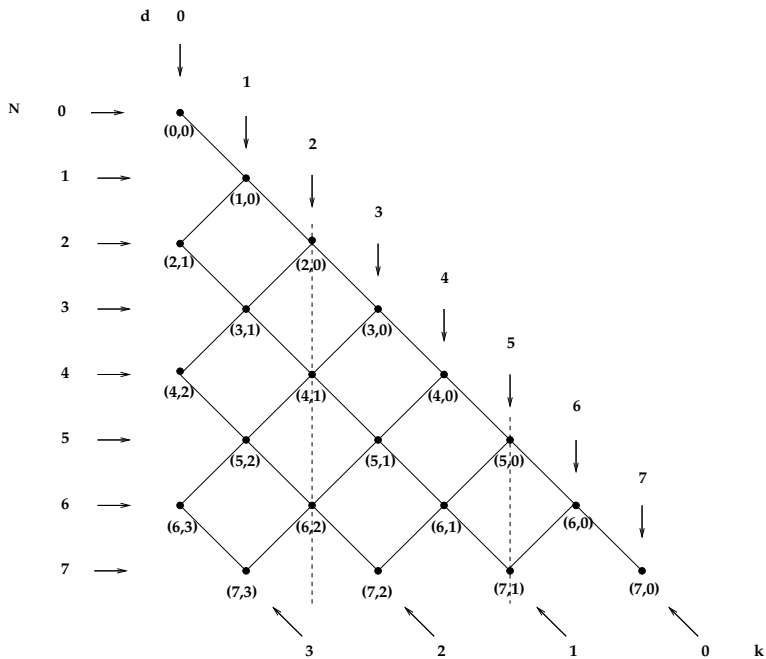}
\caption{Bratelli diagram with critical lines for $\lambda=1$}
\label{Brat2}
\end{figure}

Representations of $TL_N(\lambda)$ on the space $\mathcal{H}_N$ with
$\lambda<d$ are obtained by regarding $q$ as a formal variable with respect to
the bilinear form, i.e. complex conjugation leaves $q$ unchanged. The
operators $b_i$ project locally onto the two-site $U_q(sl_2)$ singlet but with
respect to the new bilinear form. The Temperley-Lieb parameter takes the
value
\begin{equation}
\lambda=\left [d\right ]_q=\left [2S+1\right ]_q=\sum\limits_{i=-S}^{S}q^{2i}.\label{nongenlam}
\end{equation}
The Hamiltonian (\ref{open Hamiltonian}) obtained via this type of $TL_N$
representation is then not Hermitian with respect to the usual
scalar product. The parameter $\lambda$ may now take values from the set
(\ref{zeros}). Let $i$ be the smallest integer, such that
$P_i(\lambda)=0$, then
\begin{equation}
 P_{k}(\lambda)=0\quad \Leftrightarrow\quad k=i+(i+1) n\quad\mbox{for}\; n\in\mathbb{N}.\label{zerorule}
\end{equation}
This is indicated by the dashed lines (critical lines) in figure \ref{Brat2}. In the case of such a non-generic value of $\lambda$, in general also reducible
but indecomposable $TL_N(\lambda)$-representations occur in the direct sum
decomposition of the global Hilbert space. They result from a mixing of two
generically irreducibles. This is analogous to the mixing of $U_q(sl_2)$
highest-weight representations for the $XXZ$ chain for $q$ a non-trivial root
of unity described in \cite{PasquierSaleur89}.
\newline
Let the representation $O(N,k)$ be defined as in section \ref{opensectors} as
the TL-invariant subspace obtained by starting from a vector of type
(\ref{Start}).  Some of the vectors in the construction (\ref{vasis k}) are
then no longer well defined. We consider the construction for $k=1$ and
$N=i+1$ for the condition (\ref{zerorule}). For $j<N-1$ the vector $
\mbox{v}[j;\omega_{N-2}]$ stays well defined. The vector $
\mbox{v}[N-1;\omega_{N-2}] $ stays well defined if the factor $P_{i-1}(\lambda)/P_{i}(\lambda)$ is
omitted in (\ref{vbasis}).

From equation (\ref{b action}) follows that 
\begin{equation} \tilde{ \mbox{v}}[N-1;
\omega_{N-2}]:=\left [b_{N-1}\,\mbox{v}[N-1; \omega_{N-2}]-\frac{P_{i-2}(\lambda)}{P_{i-1}(\lambda)}\mbox{v}[N-1; \omega_{N-2}]\right] \in
  \Omega_N.
\end{equation}
This means that $O(N,1)$ contains a subrepresentation of type $O(N,0)$. The
norm of $\tilde{ \mbox{v}}[N-1; \omega_{N-2}]$ with respect to the bilinear form
is zero.  Hence, there is a vector $\bar{\mbox{v}}[N-1,\omega_{N-2}]$ orthogonal
to all $\mbox{v}[j,\omega_{N-2}]$ except for $j=N-1$. From
\begin{equation}
\begin{split}
b_j\,\bar{\mbox{v}}[N-1,\omega_{N-2}]&=0,\quad\mbox{for}\quad j\le N-2,\\
b_{N-1}\,\bar{\mbox{v}}[N-1,\omega_{N-2}]&=\frac{P_1(\lambda)}{P_{N-2}(\lambda)}\mbox{b}[N-1,\omega_{N-2}],
\end{split}
\end{equation}
we find that these $N$ vectors span a reducible but indecomposable
$TL_N(\lambda)$-representation, called $I(N;1,0)$. We find the following
inclusion of subrepresentations
\begin{equation}
 O(N,0)\subset O(N,1)\subset I(N;1,0)
\end{equation}
with
\begin{equation}
 I(N,1,0)/O(N,1)\cong O(N,0).
\end{equation}
The spectrum of (\ref{open Hamiltonian}) in the space $I(N;1,0)$ is the same
as for a direct sum of $O(N,1)$ and $O(N,0)$. But compared to the generic case
the multiplicity of the ground-state energy eigenvalue of (\ref{open
  Hamiltonian}) is now given by $\mbox{dim}(\Omega_N)+\mbox{dim}(\Omega_{N-2})$.
For larger chain length the recursive definition has to be changed to
\begin{equation}
\mbox{v}[j;\omega_{N-2}]=\begin{cases}b_j\,\mathrm{v}[j-1;
 \omega_{N-2}]-\frac{P_{\overline{j-2}}(\lambda)}{P_{\overline{j-1}}(\lambda)}\mathrm{v}[j-1; \omega_{N-2}],\quad
 \mbox{for}\quad \overline{j}=i,\\b_j\,\mathrm{v}[j-1;
 \omega_{N-2}]-\mathrm{v}[j-2; \omega_{N-2}], \quad \mbox{for}\quad \overline{j}=0, \\\frac{P_{\overline{j-1}}(\lambda)}{P_{\overline{j}}(\lambda)}
 \left [b_j\,\mathrm{v}[j-1;
 \omega_{N-2}]-\frac{P_{\overline{j-2}}(\lambda)}{P_{\overline{j-1}}(\lambda)}\mathrm{v}[j-1;
 \omega_{N-2}]\right],\quad \mbox{else},
\end{cases}
\end{equation} 
with $\overline{j}:=j\, \mathrm{mod}\,(i+1)$. This construction is easily
generalized to higher $k$. For larger $N$ these indecomposable sectors induce
indecomposable sectors for higher $k$.

The multiplicity of $TL_N(\lambda)$-representations in terms of $XXZ$-$S_{\rm tot}^z$ eigenspaces is given by (\ref{sz multi}) as in the
generic case but the multiplicity of certain eigenvalues is increased (as in
the $XXZ$ chain).
 
\section{Invariant subspaces for periodic boundary conditions}\label{periodic boundaries}
Now we address the problem of determining the spectra for the periodically
closed chains.  We find that the Hilbert space $\mathcal{H}_N$ of a model with
periodically closed boundaries and $S>1/2$ can be decomposed into a direct sum
of  $PTL_N(\lambda)$-representations each isomorphic to an
$S^z_{\rm tot}$-eigenspace of an $XXZ$ chain with appropriately twisted
boundaries.

In comparison to the case of open boundaries the spectrum of our model is no
longer contained within the spectrum of a single $XXZ$ chain, 
the identification of the reference chain has to be done for each sector
separately. It follows, that the determination of the multiplicities is more
involved.

The $PTL_N(\lambda)$-representations needed here are obtained from an initial
vector $v \in \mathcal{H}_N$ with the properties

\begin{equation}
b_i\,v= \begin{cases}\lambda \, v,&\mbox{for}\quad i=2l-1 \;\mbox{with}\; 1\le l \le k,\\
0,& \mbox{for}\quad i\ge 2k+1,\end{cases}\label{altbed}
\end{equation}
and in addition
\begin{equation}
\left(b_1\,b_N\,b_{N-1}\cdots b_{2k+2}\right)\left(b_3\cdots\,b_{2k-1}\,b_{2k+1}\right)\left(b_2\cdots\,b_{2k-2}\,b_{2k}\right)\;v =a\;v\label{neubed}
\end{equation}
with some (complex) parameter $a$.  The $PTL_N(\lambda)$-representation
obtained by constructing the $PTL_N(\lambda)$-invariant subspace starting from
$v$ is determined by the two conditions (\ref{altbed}) and (\ref{neubed}) up
to isomorphism.  In contrast to the irreducible $TL(\lambda)$-representations,
the construction now depends on an additional parameter $a$. Our construction
is motivated by the Bethe ansatz.  The representation theory of the algebra
$PTL_N(\lambda)$ has been examined in \cite{PeriodicTL93} and
\cite{PeriodicTL94}, where the representations we need here occured already.

\subsection{Construction of $PTL_N(\lambda)$-representations 
for generic $\lambda$ and $\alpha=\pm\mbox{id}$
  \label{periodic sectors}}
In order to facilitate reading we restrict the construction at this point to the case 
\begin{equation}
 \alpha =\epsilon\,\mbox{id},\quad \epsilon \in\left \{1, -1\right\},\label{Alpha2}
\end{equation}
for $\alpha$ defined in (\ref{Alpha}) and complete the discussion of the general case in section \ref{generalalpha}. 
We define the space of so-called periodic reference states as
\begin{equation}
\Omega_N^p:=\{\omega_N\,\in\,\mathcal{H}_N\;:\;
b_i\,\omega_N\,=\,0\;\mbox{for all}\;1\le i\le N\}.\label{Refper}
\end{equation}
The construction becomes most clear by
 using the graphical notation (\ref{bgra}) for the operators $b_i$. 
An element of $\Omega_N^p$ will be
represented by $N$ solid dots. Starting from the vector
\begin{equation}
\mbox{\makebox[5cm][r]{$\Psi\otimes\omega_{N-2} $}
\hfill $=$ \hfill\parbox{5cm}{$~\quad\epsfig{file=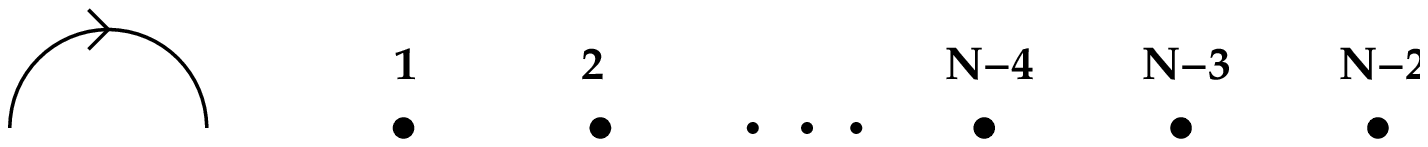,
    width= 5 cm}~ \;\quad  $}}\label{anfang}
\end{equation}
and acting on this initial state we find using (\ref{bgra})

\begin{equation}
\mbox{\makebox[5cm][r]{$b_2\;(\Psi\otimes\omega_{N-2}) $}
\hfill $=$ \hfill\parbox{5cm}{$~\quad\epsfig{file=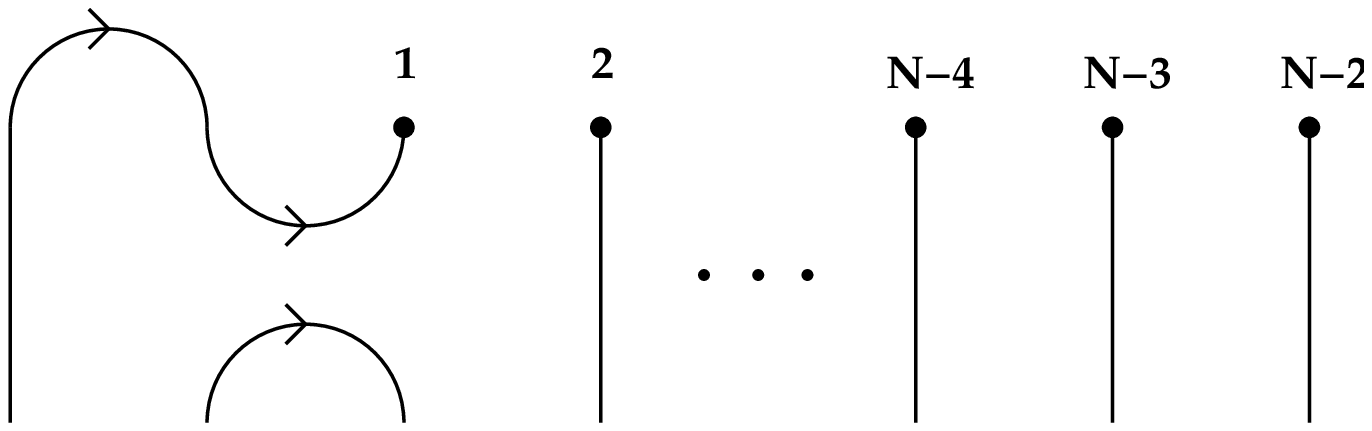,
    width= 5 cm}~ \;\quad  $}}\nonumber
\end{equation}
\begin{equation}
\mbox{\makebox[5cm][r]{$ $}
\hfill $=$ \hfill\parbox{5cm}{$~\quad\epsfig{file=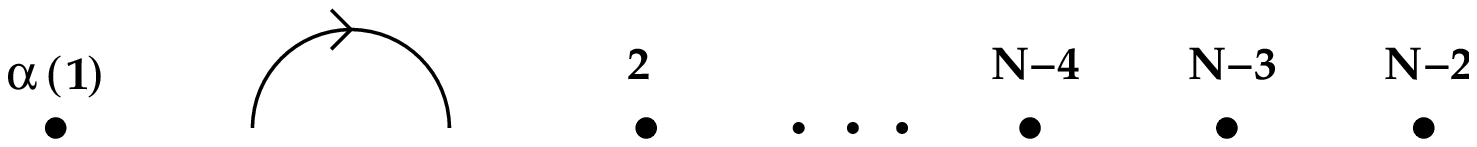,
    width= 5 cm}~ \;\quad  $}}\nonumber
\end{equation}
\begin{equation}
\mbox{\makebox[5cm][r]{$b_{N-1}\cdots\,b_3\,b_2\;(\Psi\otimes\omega_{N-2}) $}
\hfill $=$ \hfill\parbox{5cm}{$~\quad\epsfig{file=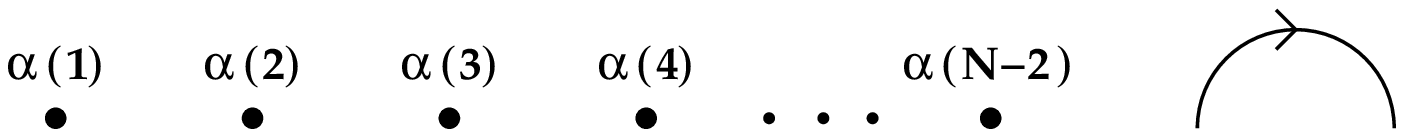,
    width= 5 cm}~ \;\quad  $}}\nonumber
\end{equation}
\begin{equation}
\mbox{\makebox[5cm][r]{$b_N\,b_{N-1}\cdots\,b_3\,b_2\;(\Psi\otimes\omega_{N-2}) $}
\hfill $=$ \hfill\parbox{5cm}{$~\quad\epsfig{file=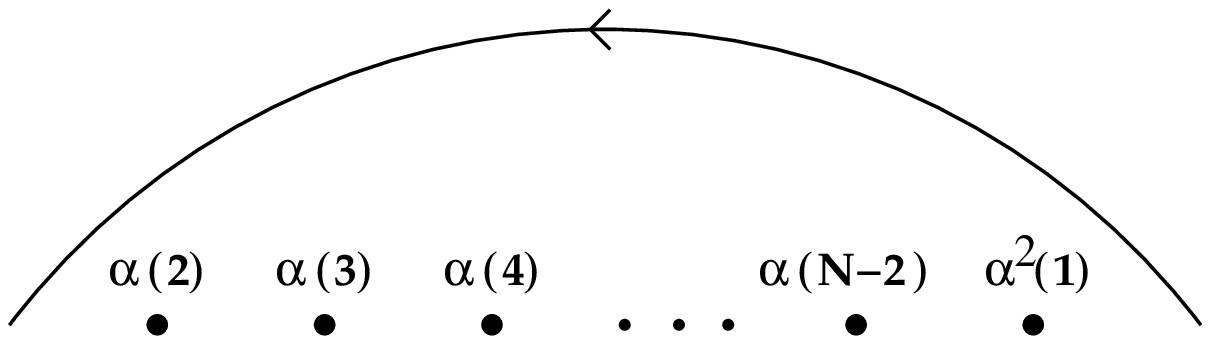,
    width= 5 cm}~ \;\quad  $}}\nonumber
\end{equation}
\begin{equation}
\mbox{\parbox{5cm}{$b_1\,b_N\,b_{N-1}\,\cdots b_2\;(\Psi\otimes\omega_{N-2})$}
\hfill $\;\:=\quad\quad$ \hfill\parbox{5cm}{$~\quad\epsfig{file=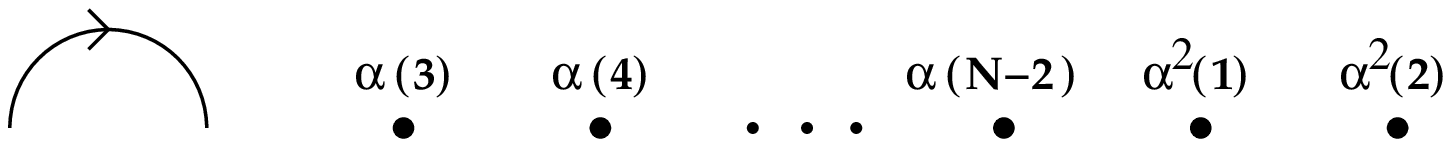,
    width= 5 cm}~ \;\quad  $}}\nonumber
\end{equation}
\begin{equation}
\mbox{\parbox{6cm}{$ $}
\hfill $\stackrel{(\ref{Alpha2})}{=} \epsilon^N$ \hfill\parbox{5cm}{$~\quad\epsfig{file=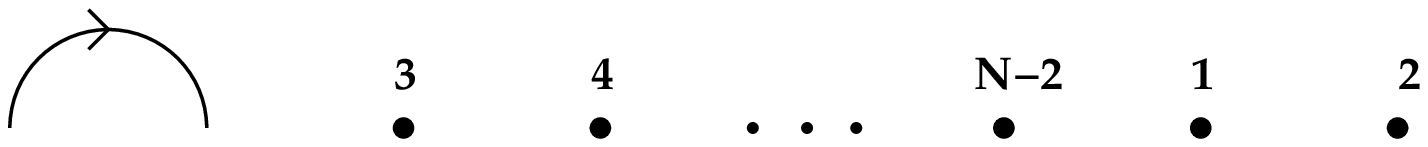,
    width= 5 cm}~ \;\quad  $}}\label{ende}
\end{equation}
Choosing $\omega_{N-2}$ to be an eigenstate of the translation operator by one site to the right $T_{N-2}$ on the $(N-2)$-fold tensor product, say $T_{N-2} \,\omega_{N-2} =e^{i\varphi}\omega_{N-2} $, (\ref{ende}) is a multiple of (\ref{anfang}).  We define the
representation $P(N,k,\omega_{N-2k}^\varphi)$ as the
$PTL_N(\lambda)$-invariant subspace constructed from the initial vector
\begin{equation}
 \mbox{b}[1,3,\cdots,2k-1; \omega^{\varphi}_{N-2k}]=\Psi^{\otimes k}\otimes\omega^{\varphi}_{N-2k} \label{kperiodic1}
\end{equation}
with 
\begin{equation}
\quad\omega^{\varphi}_{N-2k}\in\Omega^p_{N-2k}\quad\mbox{and}\quad
T_{N-2k}(\omega^{\varphi}_{N-2k})=e^{i\varphi}\omega^{\varphi}_{N-2k}.
\end{equation}
It follows that
\begin{equation}
\begin{split}
&\left(b_1\,b_N\cdots b_{2k+2}\right)\left(b_3\cdots\,b_{2k-1}\,b_{2k+1}\right)\left(b_2\cdots\,b_{2k-2}\,b_{2k}\right) \mbox{b}[1,3,\cdots,2k-1; \omega^{\varphi}_{N-2k}]\\ &=\epsilon^N e^{-2i\varphi}\; \mbox{b}[1,3,\cdots,2k-1; \omega^{\varphi}_{N-2k}]. \label{kperiodic2}
\end{split}
\end{equation}
In graphical notation, relation (\ref{kperiodic2}) means that shifting (by
acting with the TL-operators) each of the $k$ singlets by two sites to the
right and then the rightmost singlet to the initial position of the first one,
yields a multiple of the initial state (see also (\ref{ksinglets})
below). From (\ref{kperiodic2}) follows that vectors obtained by acting with
the TL-operators on the initial state (\ref{kperiodic1}), and leading to the
same distribution of singlets, are linearly dependent.
\begin{equation}
\mbox{\parbox{12cm}{$~\quad\epsfig{file=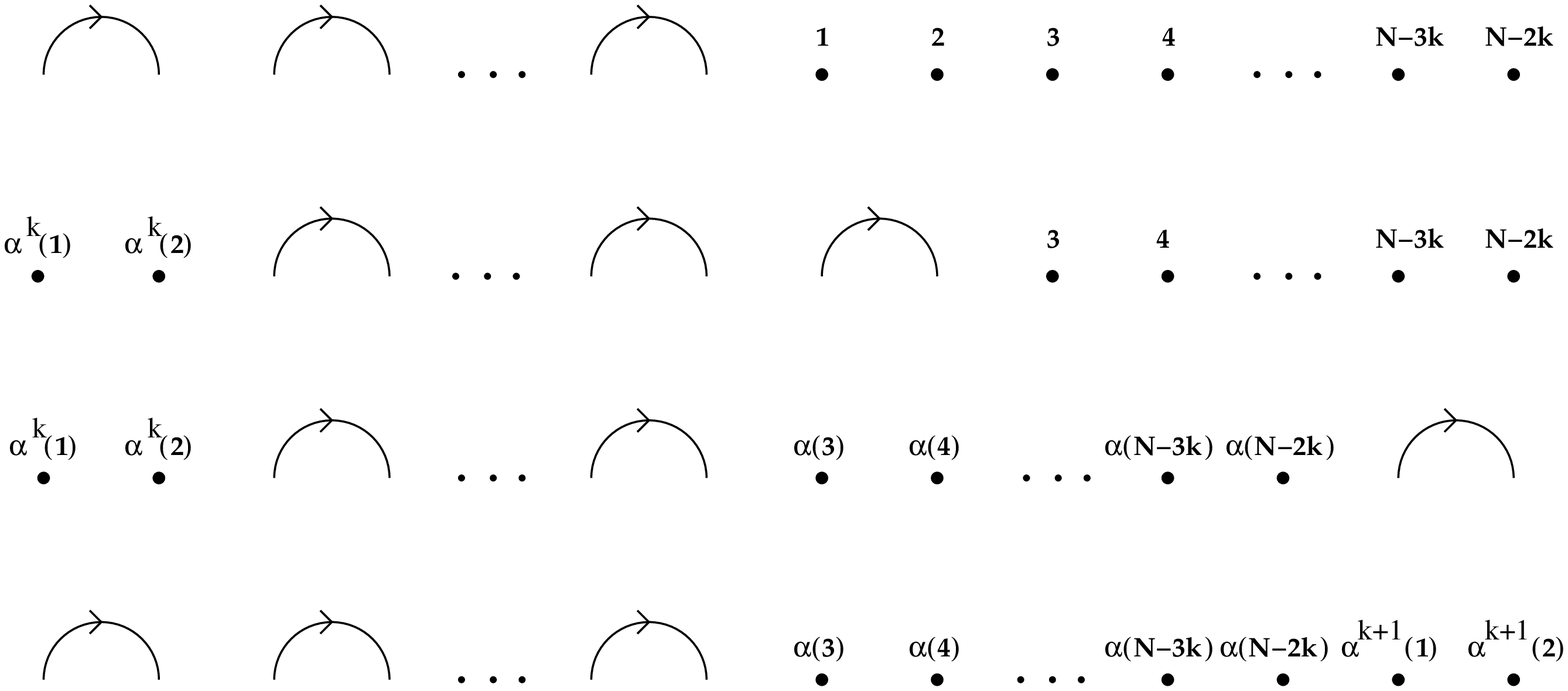, width= 10 cm}~ \;\quad  $}}\label{ksinglets}
\end{equation}
In order to construct a generating system of the $PTL_N(\lambda)$-invariant
subspace we construct the vectors
\begin{equation}
\begin{split}
 &\mbox{b}[i_1,i_2,\cdots,i_k;
\omega^{\varphi}_{N-2k}]\\&:=\left(\prod_{l=2}^{i_1}b_{l}\right)\,\left(\prod_{l=4}^{i_2}b_{l}\right)\cdots\left(\prod_{l=2k-2}^{i_{k-1}}b_{l}\right)\left(\prod_{l=2k}^{i_k}b_{l}\right)\, \mbox{b}[1,3,\cdots,2k-1; \omega^{\varphi}_{N-2k}]
\end{split}\label{gen}
\end{equation}
with the following restriction on the indices
\begin{equation}
i_l\le i_{l+1}-2\quad \mbox{for}\; l\le k-1\quad \mbox{and}\;i_k \le N
\end{equation}
which ensures that the vector defined by (\ref{gen}) is an eigenstate of
$b_i$ for $i\in\left\{i_1, i_2,\cdots i_k\right\}$. 

The operation of the local projector $\mbox{id}\otimes \left | \Psi
\right>\left<\Psi\right|\otimes \mbox{id} $ on two adjacent singlets reads in
graphical notation:
\begin{equation}
\mbox{
\parbox{3cm}{$~\epsfig{file=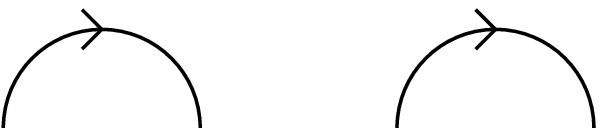, width= 2.5 cm}~$}
\hfill ${\longmapsto}$ \hfill
\parbox{3cm}{ $~\epsfig{file=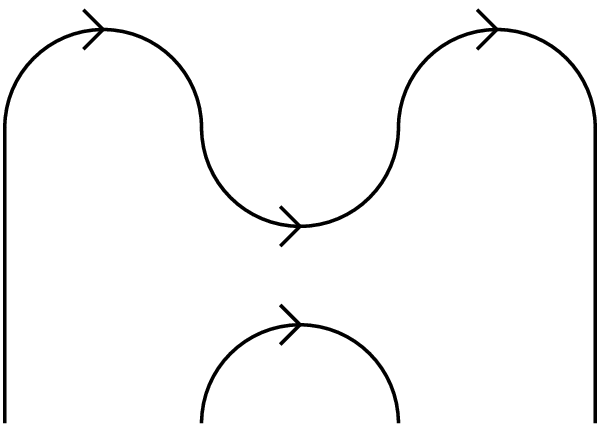, width= 2.5 cm}~$}
\hfill $=\epsilon $ \hfill
\parbox{3cm}{ $~\epsfig{file=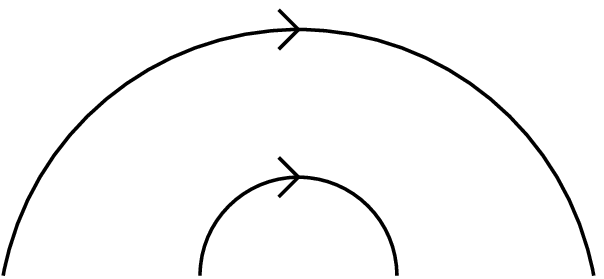, width= 2.5 cm}~$}
}  \label{psipsi}
\end{equation}
By repeated use of (\ref{psipsi}) on the vectors defined by
(\ref{gen}) every possible nesting of the $k$ singlets is
realized, yielding in total ${N\choose k}$ states. This means 
\begin{equation}
\mbox{dim}P(N,k,\omega^\varphi_{N-2k})\le {N \choose k}. \label{periodic dimension}
\end{equation}
In section \ref{periodic dec} it will be shown that equality holds.

\subsubsection{The $XXZ$ reference-model}
For the $XXZ$ representation (with $q\neq 1$) a basis of $\Omega_N^p$ is given by
\begin{equation}
\omega_N(+):=\left|+\right>^{\otimes N}\quad\mbox{and}\quad\omega_N(-):=\left|-\right>^{\otimes N}.
\end{equation}
For $\omega_N(+)$ we find for global twist angle $\phi$
\begin{equation}
b_1\,b_N\,b_{N-1}\,\cdots b_2\, \mathrm{b}[1;\omega_{N-2}(+)]=(-1)^N e^{-i\phi}\, \mbox{b}[1;\omega_{N-2}(+)]\label{condi1}
\end{equation}
and 
\begin{equation}
\begin{split}
&(b_1\,b_N\cdots b_{2k+1}\,b_{2k+2})(b_3\cdots b_{2k-1}\,b_{2k+1})(b_2\cdots
b_{2k-2}\,b_{2k}) \mathrm{b}[1,3,\cdots,2k-1;
\omega_{N-2k}(+)]\\&=(-1)^N\,e^{-i\phi}\, \mathrm{b}[1,3,\cdots,2k-1;
\omega_{N-2k}(+)]\label{condik}
\end{split}
\end{equation} 
Under the condition
\begin{equation}
\epsilon^N e^{-2i\varphi}=(-1)^N e^{-i\phi}\quad
\end{equation}
the subspace $P(N,k,\omega_{N-2k}^{\varphi})$ of a given TL quantum spin chain (\ref{Hamiltoniansp})
is isomorphic as a $PTL_N(\lambda)$-representation to the sector 
$P(N,k,\omega_{N-2k}(+))$ of the $XXZ$ chain with $\Delta=\lambda/2$ and twist angle $\phi$. Therefore the eigenvalues
of the Hamiltonians (\ref{Hamiltoniansp}) and (\ref{XXZperiodic}) coincide within these subspaces. For $q\neq 1$ the space $P(N,k,\omega_{N-2k}(\pm))$ is equal to the $S^z_{\rm tot}$ eigenspace for $s^z=\pm(N/2-k)$. 

 For the special case of the (untwisted) $XXX$ chain ($q=1$) the operator
$b_N$ can be expressed by $b_i$ with $1\le i \le N-1$ as follows 
\begin{equation}
b_N=(\mbox{id}-b_{N-1})\cdots (\mbox{id}-b_{3})(\mbox{id}-b_{2}) b_1 (\mbox{id}-b_{2})(\mbox{id}-b_{3})\cdots(\mbox{id}-b_{N-1}),\label{braidie}
\end{equation} 
meaning that every $TL_N(\lambda)$-representation $O(N,k)$ is already closed under operation of $b_N$.

\subsection{Decomposition of $PTL_N(\lambda)$-representations \label{periodic dec}}
$TL_N(\lambda)$ is a subalgebra of $PTL_N(\lambda)$, so every subspace
$P(N,k,\omega^\varphi_{N-2k})$ is a $TL_N(\lambda)$-representation by omitting
the operator $b_N$. It follows that $P(N,k,\omega^\varphi_{N-2k})$ decomposes
into a direct sum of irreducible $TL_N(\lambda)$-representations in the
generic case. We find
\begin{equation}
P(N,k,\omega^\varphi_{N-2k})\downarrow_{TL_N(\lambda)}\cong
\oplus_{l=0}^{k} O(N,l).\label{per decomposition}
\end{equation}
For the proof it suffices to give the initial vectors generating the
$TL_N(\lambda)$-representations on the \rhs of (\ref{per
  decomposition}).
\newline
For a vector $\omega^\varphi_{N-2k}\in\Omega^p_{N-2k}$ we construct a sequence
of vectors
\begin{equation}
\omega_{N-2k+2l}(\omega^\varphi_{N-2k})\in\Omega_{N-2k+2l},\quad  l=1, \dots, k.\label{omegas}
\end{equation}
For $l=1$ it follows from the construction in the previous chapters that
\begin{equation}
\omega_{N-2k+2}:=T^{-1} \mathrm{b}[1;\omega^{\varphi}_{N-2k}]-\sum_{i=1}^{N-2k+1}C( \mathrm{v}[i])\, \mathrm{v}[i;\omega^\varphi_{N-2k}]\label{om1}
\end{equation}
with $T$ the $(N-2k+2)$-site translation operator and coefficients
\begin{equation}
C( \mathrm{v}[i])=\begin{cases}(-1)^{\mathrm{i}}\epsilon^{N-2k+2}
  e^{ i\varphi}\frac{1}{P_1(\lambda)},\quad
  &
  \text{for}\quad i<N-2k+1,\\
  (-1)^{i}\epsilon^{N-2k+2} e^{ i\varphi}\frac{1}{P_1(\lambda)}-e^{ -i\varphi}\frac{P_{N-2k}(\lambda)}{P_1(\lambda)},\quad
  & \text{for}\quad i=N-2k+1,\end{cases}
\end{equation}
is an element of $\Omega_{N-2k+2}$. For $l\ge 2$ with $\tilde{N}:=N-2(k-l)$ we
find recursively
\begin{equation}
\omega_{\tilde{N}}:=T^{-1}\mbox{b}[1; \omega_{\tilde{N}-2}]-\sum_{i=1}^{\tilde{N}-1}
C_i\, \mbox{v}[i; \omega_{\tilde{N}-2}]-\sum_{i=1}^{\tilde{N}-2}
\tilde{C}_i\, \mbox{v}[i,\tilde{N}-1; \omega_{\tilde{N}-4}]\label{oml}
\end{equation}
with the coefficients
\begin{equation}
C_i=\begin{cases}(-1)^{l+i-1}\epsilon^{(l-1)\tilde{N}}\left((-1)^{\tilde{N}}
    e^{-i\varphi}\frac{P_{l-2}(\lambda)}{P_1(\lambda) P_{\tilde{N}-3}(\lambda)}+\epsilon^{\tilde{N}}
    e^{i\varphi}\frac{P_{\tilde{N}-l-2}(\lambda)}{P_1(\lambda) P_{\tilde{N}-3}(\lambda)}\right), &\text{for}\quad
    i<\tilde{N}-1,\\ 
 (-1)^{l+i-1}\epsilon^{(l-1)\tilde{N}}\left((-1)^{\tilde{N}}
    e^{-i\varphi}\frac{P_{\tilde{N}-l-1}(\lambda)}{P_1(\lambda)}+\epsilon^{\tilde{N}} e^{
    i\varphi}\frac{P_{l-1}(\lambda)}{P_1(\lambda)}\right),& 
    \text{for}\quad i=\tilde{N}-1,\end{cases}  
\end{equation}
and
\begin{equation}
\tilde{C}_i=(-1)^{i}\frac{P_{\tilde{N}-l-2}(\lambda)\,P_{l-2}(\lambda)}{P_1^2(\lambda)}\frac{D_{\tilde{N}-2}^{\varphi,\epsilon}(\lambda)}{P_{\tilde{N}-3}(\lambda)}.\nonumber
\end{equation} 
The polynomials $D_k$ are defined by
\begin{equation}
 D_k^{\varphi,\epsilon}(x):=P_{k}(x)-P_{k-2}(x)-(-\epsilon)^{k}(e^{2i\varphi}+e^{-2i\varphi})\quad\mbox{for}\quad k\ge 2.   
\end{equation}
For the square of the norm one finds 
\begin{equation} 
\begin{split}
\left<\omega_{\tilde{N}}\right|\left.\omega_{\tilde{N}}\right>=&\frac{P_1(\lambda) P_2(\lambda)\cdots
  P_{l-1}(\lambda)}{P_{\tilde{N}-1}(\lambda) P_{\tilde{N}-2}(\lambda)\cdots
  P_{\tilde{N}-l}(\lambda)}D_{\tilde{N}}^{\varphi,\epsilon}(\lambda) D_{\tilde{N}-2}^{\varphi,\epsilon}(\lambda)\cdots
  D_{\tilde{N}-2(l-1)}^{\varphi,\epsilon}(\lambda)\\ =&\prod_{i=1}^l\frac{P_{i-1}(\lambda)}{P_{\tilde{N}-i}(\lambda)}D_{\tilde{N}-2(i-1)}(\lambda).\label{normomega}
  \end{split}
\end{equation}
For the Temperley-Lieb parameter $\lambda$ in the semisimple regime
and $\varphi \in \mathbb{R}$ we find $D_k(\lambda)\neq 0$ for all
$k$. \newline
From the construction of the vectors in the sections \ref{opensectors}
and \ref{periodic sectors} it can be checked that (\ref{omegas}) holds
for (\ref{om1}) and (\ref{oml}). It follows that 
\begin{equation}
\mbox{b}[1, \dots, 2l-1; \omega_{N-2k+2l}(\omega^\varphi_{N-2k})]
\end{equation}
yields a $TL_N(\lambda)$-representation $O(N, k-l)$. With the upper
threshold for $\mbox{dim}(P(N,k;\omega^\varphi_{N-2k}))$ found in
section \ref{periodic sectors} equation (\ref{per decomposition}) follows. 
The operator $b_N$ acts on the vector $\omega_N$ from (\ref{omegas}) as 
\begin{equation} 
b_N\,\omega_N(\omega^\varphi_{N-2k})=\frac{P_{k-1}(\lambda)\,P_{N-k-1}(\lambda)}{P_{N-1}(\lambda)\,P_{N-2}(\lambda)}D^{\varphi,\epsilon}_{N}(\lambda)\;T^{-1}\mbox{b}[1; \omega_{N-2}]. \label{bNaction}
\end{equation} 
This shows $\omega_N \notin \Omega_N^p$. The $PTL_N(\lambda)$-representations
$P(N,k,\omega^\varphi_{N-2k})$ are generically irreducible.
\subsection{The sector with $k=N / 2$ \label{sing}} 
For even values of the chain length the subspace $P(N,N/2)$ is of
special importance because it yields the eigenvector of largest
absolute eigenvalue of (\ref{Hamiltoniansp}). For  
\begin{equation} 
 \mbox{b}[1,3,\cdots,N-1]=\left|\Psi\right>^{\otimes \frac{N}{2}} 
\end{equation} 
we find   
\begin{equation} 
\left(b_3\cdots\,b_{2k-1}\,b_{1}\right)\left(b_2\cdots\,b_{2k-2}\,b_{2k}\right)
\mbox{b}[1,3,\cdots,N-1]=\left[\mbox{tr}\left(\alpha^{\frac{N}{2}}\right)\right]^2\; \mbox{b}[1,3,\cdots,N-1] \label{Dreherk}
\end{equation} 
We give the proof in graphical notation. (To keep the graphical presentation
simple we consider $N=6$)
\begin{equation} 
\mbox{\parbox{4 cm}{ $~\epsfig{file=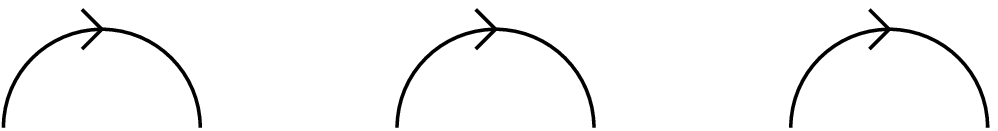, width= 4 cm}~ $} 
\hfill $\quad\stackrel{b_2\,b_4\,b_6}{\longrightarrow}\quad$ \hfill  
\parbox{5 cm}{ $~\epsfig{file=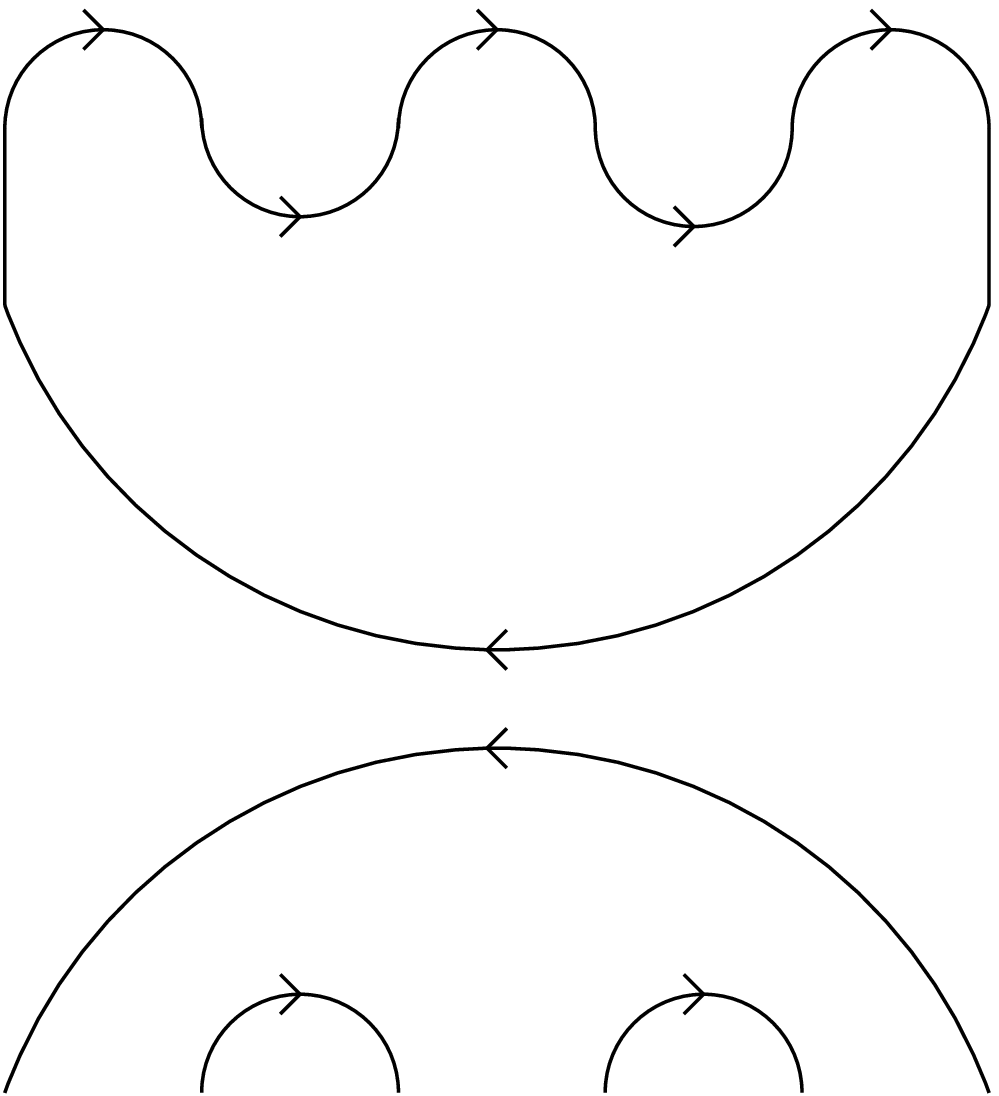, width= 4 cm}~ $}}
\label{Halbzeit} 
\end{equation} 
Acting with $b_3\,b_5\,b_1$ on the \rhs 
of (\ref{Halbzeit}) shows equation (\ref{Dreherk}). 
The corresponding twist angle is given by  
\begin{equation} 
 \varphi=i\,\ln\left(\left
 |\mbox{tr}\left(\alpha^{\frac{N}{2}}\right)\right |+\sqrt{\left[\mbox{tr}\left(\alpha^{\frac{N}{2}}\right)\right]^2-4}\right)-i\,\ln(2). 
\label{Twisty} 
\end{equation} 
In the special case of  $\left|\Psi\right>$ defined by
(\ref{Uqsinglet}) for $q=1$ we have 
$P_2\,\left|\Psi\right>=\pm \left|\Psi\right>$ for $P_2$ the two-site
permutation operator. We call this an isotropic singlet. In this case the decomposition formula reduces
to
\begin{equation}
P(N,N/2)\downarrow_{TL_N(\lambda)}\cong O(N,N/2).
\end{equation}
For $q\neq 1$ we find along the lines of section \ref{periodic dec}
\begin{equation} 
P(N,N/2)\downarrow_{TL_N(\lambda)}\cong
\oplus_{l=0}^{N/2} O(N,l).
\end{equation}

\subsection{Dimension of $\Omega_N^p$}
    \label{dimension omegap}
The dimension of the space $\Omega_N^p$ of periodic reference states, i.e. the
multiplicity of the trivial representation of $PTL_N(\lambda)$ in the space
$\mathcal{H}_N$ for $\mbox{dim}(h)=d$ is given by
\begin{equation}
\mbox{dim}(\Omega_N^p)=\mbox{dim}(\Omega_N)-\mbox{dim}(\Omega_{N-2})\quad\mbox{for}\;N>2.\label{rec dim}
\end{equation}
Proof: Consider the map
\begin{equation}
b_N\;:\;\Omega_N \longrightarrow \left |\Psi\right >\otimes
\Omega_{N-2} \;\subset h_N\otimes h_1\otimes \cdots \otimes
h_{N-1}, \label{Perabb} 
\end{equation}
where $\mathcal{H}_N$ is considered as $ h_N\otimes h_1\otimes \cdots \otimes
h_{N-1}$.  
We show the surjectivity of the map (\ref{Perabb}) by induction
over the chain length $N$. For $N=1$ we have
$\mbox{dim}(\Omega_1^p)=\mbox{dim}(h)$ because $\Omega^p_1=\Omega_1=h$. For
$N=2$ we find in the case of an anisotropic singlet
\begin{equation}
\mbox{dim}(\Omega_2^p)=d^2-2 
\end{equation}
because the eigenspaces of $b_1$ and $b_2$ are distinct and
one-dimensional. Suppose equation (\ref{rec dim}) holds for all $N'<N$. From
the induction hypothesis it follows that
\begin{equation}
\sum_{k=0}^{\left[(N-2)/2\right]}\mbox{dim}\Omega^p_{N-2-2k}=\mathrm{dim}(\Omega_{N-2}).\label{summ} 
\end{equation}
From the decomposition rule (\ref{per decomposition}) and equation (\ref{bNaction})
we know that every representation $P(N,k)$ with $1\le k\le \left[N/2\right]$
contains an element of $\Omega_N$ which is not an element of
$\Omega^p_N$. The number of these independent states is equal to the
lhs of (\ref{summ}). On the space spanned by these states $b_N$
acts injectively. Hence the dimension of the image of $b_N$ is larger than the rhs of
(\ref{summ}), which proves surjectivity of $b_N$ as in
(\ref{Perabb}). 
The dimension of the space of periodic reference states is then given by
\begin{equation}
\mbox{dim}(\Omega_N^p)(d)=\left ( \frac{d}{2} +\frac{ \sqrt{d^2-4}}{2}
\right )^N + \left ( \frac{d}{2} - \frac{\sqrt{d^2-4}}{2} \right
)^N.\label{dimension ref}
\end{equation}
An exception of equation (\ref{dimension ref}) occurs in case of an isotropic
singlet. For the $\lambda=2$ XXX chain we have $\Omega_N^p=\Omega_N$ because
of (\ref{braidie}). For the other isotropic singlets we find
\begin{equation}
\mbox{dim}(\Omega_2^p)(d)= d^2-1\quad \mbox{instead of}\quad d^2-2,
\end{equation}
but for $N\ge 2$ equation (\ref{dimension ref}) holds again because the higher
dimension of $\Omega_2^p$ compensates for the fact that the sector for
$k=N/2$ does not contain an open reference state in this case.

\subsection{The non-generic case\label{nongenericP}}
For the representations discussed in section (\ref{nongenericO}) for certain
values of $\lambda<d=2S+1$ the direct sum decomposition of the global Hilbert
space contains reducible but indecomposable representations obtained from the
mixing of generically irreducibles. Let $\lambda$ be generic with respect to
the algebra $TL_N(\lambda)$. In case that $\lambda$ is a zero of the
polynomial $D_N^{\varphi,\epsilon}$, equations (\ref{normomega}) and (\ref{bNaction}) show that
the vector $\omega_N$ constructed in $P(N,1,\omega^p_{N-2})$ belongs to
$\Omega^p_N$ and belongs to its own orthogonal complement with respect
to the bilinear form of section \ref{nongenericO}. There then exists a vector $\tilde{\omega}_N$ with
\begin{equation}
\tilde{\omega}_N \in \Omega_N,\quad b_N\,\tilde{\omega}_N=T_N^{-1}\mbox{b}[1; \omega^p_{N-2}]
\end{equation}
and the multiplicity of the ground state eigenvalue is increased. For chain
length $N+2(k-1)$ a mixing of a $k$-singlet and a $(k-1)$-singlet sector is
induced.  The positions of the zeros of the polynomials
$D^{\varphi,\epsilon}_k$ depend on the value of $\varphi$, we skip a detailed
analysis of the situation.
\newline For $\lambda$ nongeneric with respect to $TL_N(\lambda)$ the summands
in the decomposition formula (\ref{per decomposition}) mix as described in
section (\ref{nongenericO}). The existence of a $PTL_N(\lambda)$-invariant
subspace depends again on the value of $\varphi$.
\subsection{The spectrum of the translation operator in the space $\Omega_N^p$}\label{Winkel}
The space of periodic reference states $\Omega_N^p$ is an eigenspace of the
Hamiltonian $H^p$ defined by (\ref{Hamiltoniansp}) and the translation operator
$T_N$ commutes with $H^p$, which means that $T_N$ is diagonalisable within the
space $\Omega_N^p$.

\subsubsection{ Eigenvalues and multiplicities in the global Hilbert space}\label{glohil}
To determine the eigenspectrum of the translation operator $T_N$ on
the global Hilbert space $\mathcal{H}_N$ of an $N$-site spin-$S$ chain we take
for the local Hilbert space the basis $B_S$ (see (\ref{spinbasis})).  The
cyclic group $C_N$ generated by $T_N$ acts on the basis
\begin{equation}
B^{ N}:= \left\{\, \left| M_1 \right >\otimes \left| M_2 \right >\otimes \cdots \otimes \left| M_N \right >,\quad\quad \left| M_i \right >\in B_S \right\}
\end{equation}
of $\mathcal{H}_N$.  It follows, that the set $B^{N}$ has a partition of
$C_N$-orbits. The length $p$ of a given orbit is the period of each element of
this orbit, i.e. $p$ is the smallest integer greater than zero such that
\begin{equation}
\left (T_N\right )^p (\left| M_1 \right >\otimes \left| M_2 \right >\otimes \cdots \otimes \left| M_N \right >)=(\left| M_1 \right >\otimes \left| M_2 \right >\otimes \cdots \otimes \left| M_N \right >).
\end{equation}
There are elements with period $p$ in $B^{N}$ iff $p$ divides $N$, furthermore
the number of such elements is independent of $N$. This means that defining
$\sigma\left(p\right)$ as the number of elements with period $p$ in the set
$B^{p}$, the dimension of the global Hilbert space can be written as
 \begin{equation}
\sum_{p\,|\,N}\sigma(p)=\mbox{dim}(\mathcal{H}_N).\label{sump}
\end{equation} 
Solving equation (\ref{sump}) yields
\begin{equation}
\sigma(N)=\sum_{r\,|\,N}\mu\left(\frac{N}{r}\right)\;\mbox{dim}(\mathcal{H}_r).\label{sigma}
\end{equation}
Where $\mu$ is the M\"obius function defined by
\begin{equation*}
\mu(d):=\begin{cases}
1,&\text{for $d=1$},\\
(-1)^s,&\text{if $d$ is the product of $s$ distinct primes},\\
0,&\text{else}.\end{cases} 
\end{equation*}
So for every divisor $p$ of $N$ there are $\sigma(p)/p$ many multiplets
 \begin{equation}
M_p:=\left\{e^{i\frac{2\pi l}{p}}\;:\; 0\le l\le p-1\right\}\label{multi}
\end{equation} 
of $T_N$-eigenvalues within the global Hilbert space $\mathcal{H}_N$.

\subsubsection{Multiplicities in the space $\Omega_N^p$}
The coefficients of the eigenvectors of $T_N$ within the space $\Omega_N^p$ depend continuously on $q$
for a representation defined via (\ref{Uqsinglet}), while the corresponding
eigenvalues of $T_N$ stay constant. From section \ref{dimension omegap} it
is known, that the dimension of the space $\Omega_N^p$ is independent of
$q$. It follows that the multiplicity of a given $T_N$-eigenvalue in the space
of periodic reference states is independent of q. To determine the
multiplicities we examine the limit $q\rightarrow \infty$. In this case each
operator $b_i$ projects locally on the vector
$\left|-S\right>\otimes\left|S\right>$ which means that for this special case
the set
\begin{equation}
\tilde{B}^{N}:=\left\{\, \left| M_1 \right >\otimes \left| M_2 \right >\otimes \cdots
  \otimes \left| M_N \right >,\quad
  M_{i+1}-M_i\neq 2S,\quad M_1-M_N\neq 2S
\right\}\subset B^{N}
\end{equation}
 provides a basis of $\Omega_N^p$. Along the lines of section
 \ref{glohil} we find
\begin{equation}
 \sum\limits_{p|N}\tilde{\sigma}(p)=\mbox{dim}(\Omega^p_N)\label{sump2}
\end{equation}
and
\begin{equation}
\tilde{\sigma}(N)=\sum_{r\,|\,N}\mu\left(\frac{N}{r}\right)\;\mbox{dim}(\Omega^p_r).\label{sigma2}
\end{equation}
Here $\tilde{\sigma}(N)$ is the number of elements of period $N$ in
 the set $\tilde{B}^{N}$. 
 For $l\in\mathbb{N}$ with  $0\le l \le N-1 $ the momentum
\begin{equation}
\frac{2\pi l}{N}=2\pi\frac{r}{N}\frac{l}{r}\qquad\mbox{for} \quad r\,|\,\mbox{gcd}(l,N)\nonumber
\end{equation} 
occurs in every orbit with period $N/r$. Thus we find for the multiplicity of
this momentum in the space of reference states
\begin{equation}
 M\left (\frac{2\pi l}{N},\Omega^p_N\right )=\sum\limits_{r |(N,l)}\frac{r}{N}\tilde{\sigma}(N/r).\label{mult2}
\end{equation}

\subsection{ General twisted boundaries : $\alpha \neq \pm id$} \label{generalalpha}
Relation (\ref{ksinglets}) shows that in order to obtain a representation of
the desired type for $\alpha\neq \pm id$ the vector $\omega_{N-2k}$ has to lie
in the simultaneous kernel of the operators $b_1, \dotsc, b_{N-2k-1}$ and
$b^{\alpha k}_{N-2k}$ with the latter defined by
\begin{equation}
b^{\alpha k}|_{h_{N-2k}\otimes h_{1}}= \left(id\otimes
  \alpha^{-k}\right)\circ \left|\Psi\right>\left<\Psi\right|
  \circ \left(id\otimes \alpha^{k}\right) \nonumber
\end{equation}
and as identity elsewhere. Set
\begin{equation}
\tilde{\Omega}_{N-2k}^p:=\left\lbrace \omega\in \mathcal{H}_{N-2k}:\quad b_i\,\omega=0,\,i<N-2k;\; b^{\alpha k}_{N-2k}\,\omega=0\right\rbrace .
 \end{equation}
Furthermore $\omega_{N-2k}$ has to be an eigenstate of the translation
operator followed by a twist at the last position:
\begin{equation}
 T_{N-2k}^{\alpha^k}:=\left(id^{\otimes N-2k-1}\otimes \alpha^{k}\right)\circ T_{N-2k}.
\end{equation}
The map $\alpha^k$ is diagonal with respect to the basis $B_S$ of
$S^z_{\rm tot}$-eigenstates (see (\ref{alphaformula}))
 \begin{equation}
\alpha^k\; :\;\left|M\right>\mapsto
(-1)^{2 S k}e^{-2i\phi M k}\left|M\right>.
\end{equation}
For the construction of the invariant subspaces $P(N,k,\omega_{N-2k})$ the
vectors $\omega_{N-2k}$ have to be simultaneous eigenstates of $T_{N-2k}^{\alpha^k}$ and $S^z_{\rm tot}$. The
effective twist angle then depends on both the momentum and the
$S^z_{\rm tot}$ eigenvalue of $\omega_{N-2k}$.

An element of $B_S$ with period $p$ and $S_{\rm tot}^z$-eigenvalue $s^z (=M_1+...+M_N)$ satisfies
\begin{equation}
\left (T_N^\alpha\right )^p (\left| M_1 \right >\otimes \left| M_2 \right >\otimes \cdots \otimes \left| M_N \right >)=\left (\epsilon\, e^{- i \frac{2 \phi s^z}{N}}\right)^p(\left| M_1 \right >\otimes \left| M_2 \right >\otimes \cdots \otimes \left| M_N \right >).\nonumber
\end{equation}
The corresponding orbit yields the $T_N^\alpha$ eigenvalues 
\begin{equation}
 \left\{\epsilon \,e^{-i\frac{2\phi s^z}{N}} e^{i\frac{2\pi l}{p}}:\quad 0\le l\le p-1\right\}.
\end{equation}
For $T_N^{\alpha ^k}$ the eigenvalues obtained for period $p$ and
$S^z_{\rm tot}$-eigenvalue $s^z$ are
\begin{equation}
 \left\{ \epsilon^k \;e^{i\left(\frac{2\pi
 l}{p}-\frac{2\phi s^z k}{N}\right)}:\quad 0\le l\le p-1\right\}.\nonumber
\end{equation}
To obtain the correct multiplicities we have to refine the diagonalisation of
the translation operator by distinction of $S^z_{\rm tot}$ eigenvalues.
\newline
For the number of elements with $S^z_{\rm tot}$ eigenvalue $s^z$ in the set
$\tilde{B}^N$, denoted by $\tilde{\nu}(s^z;N)$, we find
\begin{equation}
\tilde{\nu}(s^z;N)=\sum_{ \frac{N}{p}|(N, s^z)}\tilde{\sigma}\left(p, s^z\frac{p}{N}\right).
\end{equation}
Here $\tilde{\sigma}(p, s^z)$ is defined as the number of elements with period
$p$ and $S^z_{\rm tot}$-eigenvalue $s^z$ in $\tilde{B}^{N}$.  In this context
$d\in \mathbb{N}$ is a divisor of $s^z$ iff $s^z / d$ is an admissible
eigenvalue of $S^z_{\rm tot}$.  We find
\begin{equation}
\tilde{\sigma}(N,s^z)=\sum_{d\,|(N, s^z)}\mu\left(\frac{(N,
    s^z)}{d}\right)\;\nu\left(\frac{s^z}{d};\frac{N}{d}\right).
\end{equation}
The multiplicity of the momentum 
\begin{equation}
\frac{2\pi l-2 \phi s^z}{N}\qquad \mbox{with}\;\;0\le l \le N-1,\nonumber
\end{equation}
in the spectrum of $T^\alpha$ acting on the space $\Omega_N$ is given by
\begin{equation}
M\left( \frac{ 2\pi l-2 \phi s^z }{N}; \Omega^p_N \right )=\sum_{r\,|\,(N, l)}\frac{r}{N}\tilde{\sigma}\left(\frac{N}{r}, \frac{s^z}{r}\right).
\end{equation}

\section{Applications to the thermodynamics of quantum spin chains}\label{applications}
In this section we employ our mathematical results on irreducible 
representations of the Temperley-Lieb algebra to the study of physical
properties of quantum spin chains. The ordinary Temperley-Lieb equivalence of
TL-models applies to the case of open boundary conditions
\cite{BaxterBook82,PMBook91}: the eigenvalues, but not the multiplicities, can
be calculated by comparison to the spin-1/2 $XXZ$ chain. Below, we first point
out physical quantities that have to be studied on a lattice with periodic
boundary conditions, and second we deal with quantities for which the proper
treatment of multiplicities matters.

Of particular interest are the ground-state properties of a system. Usually, a
many body system is gapped and may show long-range order, or it is gapless and
exhibits critical behaviour. In the gapped case, the study of the system on a
lattice with open boundary conditions is sufficient to find results by a
mapping to the spin-1/2 $XXZ$ chain on an open lattice. In the gapless case,
i.e. for a critical system it is extremely more profitable to study the model
on a lattice with periodic boundary conditions. For this case, there exist
scaling relations of conformal field theory connecting the scaling dimensions
to the low-lying energy data of the Hamiltonian. Clearly, the `weak'
Temperley-Lieb equivalence of TL-models with open boundaries is not
applicable. An erroneous application of this kind would result into (wrong)
critical indices identical to those of the $XXZ$ chain.

In fact, the correct Temperley-Lieb equivalence is the one established in
section~\ref{periodic boundaries} relating the given TL-model on a lattice with
periodic boundary conditions to the $XXZ$ chain with suitable twist
angles. For the latter case, the Bethe ansatz equations are known, see
(\ref{BAnsatz1}). Similar results were derived, however by different
reasoning, for the $RSOS$ models and for the quantum version of the Potts
model in \cite{BazhanovReshetikhin89,AlcarazBBBQ1987}.

In this work, we do not further consider such applications as the spin chains
introduced above have gapped excitations for zero magnetic fields and hence do
not show critical properties. The ground-state energy, excitations and the
excitation gap were calculated in
\cite{Parkinson88,BarberBatchelor89,KlumperBiq89,KlumperBiq90a,KlumperBiq90b},
the correlation length was treated in
\cite{KlumperBiq89,KlumperBiq90a,KlumperBiq90b,SoerensenYoung90}.  Most of
these calculations were carried out for the spin-1 quantum chain which can be
understood as a special point in the strong coupling limit of an ionic Hubbard
model \cite{BatistaAligia2004}. In fact, for vanishing external fields the
systems are dimerized. For an odd number of sites, the ground-state is just the
lowest-lying state in a continuum of one-particle states
\cite{Albertini2000}. The situation changes drastically if anisotropies are
introduced \cite{AlcarazMalvezzi92} or an external magnetic field exceeding
the spectral gap. These cases will be studied elsewhere.

In this work, we are more concerned with the thermodynamical properties of the
quantum spin chains of TL-type. The main physical result of these applications
is a `Temperley-Lieb equivalence at finite temperature and finite magnetic
field'.

The starting point of thermodynamical studies is the so-called partition
function $Z_N$ which in our case reads
\begin{equation}
 Z_N(T,h)={\rm Tr}\, e^{-\beta (H-h M)}
\end{equation}
where $H$ is the Hamiltonian, $h$ the magnetic field, $\beta:=1/T$ the
reciprocal of the temperature $T$ and $M=S^z_{\rm tot}$ the magnetization
operator on a system of size $N$. Since we are interested in $Z_N$ and related
quantities in the thermodynamic limit $N\to\infty$, the choice of boundary
conditions is not expected to matter.

The spectrum of a given TL-Hamiltonian (for vanishing external field) and that
of the related $XXZ$ chain in equivalent $k$-sectors (our short hand for the
representations $O(N,k)$ resp.  $P(N,k)$) are identical. However, only for the
$XXZ$ chain the multiplicity of the considered representation is simple and
identical to one or two. For the other systems, the multiplicities were derived in
sections~\ref{open boundaries} and \ref{periodic boundaries}. Asymptotically,
for large $N$ and $N-2k$, the multiplicity of the $k$-sector is $z^{N-2k}$
with the `fugacity' $z$ defined by
\begin{equation}
z:=\left({\frac{d + \sqrt{d^2-4}}2}\right).
\end{equation}
For the $XXZ$ chain, the quantum number $k$ is the number of flipped spins
with respect to the ferromagnetic state. Therefore the eigenvalues of the
magnetization operator are $M=N/2-k$.

For computing the partition function, we sum over all sectors and within each
one over all energies
\begin{equation}
Z(T,h=0)\simeq\sum_{k=0}^Nz^{N-2k}\sum_{{\rm all} E_k} e^{-\beta E_k},
\end{equation}
which gives the grand-canonical partition function of the $XXZ$ reference
model with a non-vanishing magnetic field
\begin{equation}
Z(T,h=0)={\rm
  Tr}\,
e^{-\beta(H_{XXZ}-(T\ln z) 2M)}=Z_{XXZ}(T,2T\ln z).
\label{partition0}
\end{equation}
Note, this line of arguments applies only in the case of vanishing external
field for the TL-Hamiltonian. If we include a finite field, all energy
eigenvalues in a $k$-sector will be shifted by the same Zeeman term, but
equivalent, however different sectors will have different shifts. The reason
lies in the construction of the $k$-sectors: the states of the space
$\Omega_{N-2k}^{(p)}$ with different magnetizations enter.

There is, however, an alternative method for the calculation of the partition
function avoiding the explicit study of the Hamiltonian, see for instance
\cite{KlumperTherm93} and references therein. The alternative employs a
mapping of the quantum chain of length $L$ to a classical 2-dimensional system
of size $L\times N$, where $N$ is usually referred to as Trotter number which
has to be sent to infinity. Subsequently, an analysis of just the largest
eigenvalue of the quantum transfer matrix (QTM, i.e.~the transfer matrix
describing the evolution in chain direction) yields the partition
function. The temperature and magnetic field of the quantum chain appear as
staggering parameters of the local spectral parameters and as twist angle of
the periodic boundary conditions of the quantum transfer matrix,
respectively. The largest eigenvalue of the QTM lies in the $N/2$-sector,
i.e.~in the unique copy of $P(N,N/2)$.

The computational strategy is clearcut. We denote temperature and magnetic
field for the TL-models of section~\ref{TL-models} by $T$ and $h$,
respectively. The $N/2$-sector is characterized by the twist angle $\varphi$,
or equivalently by the number corresponding to the `loop' depicted in
(\ref{Halbzeit}). This is the trace of the boundary operator
\begin{equation}
{\rm Tr}\,\exp(\beta h M)=\frac{\sinh\left( ( S+\frac 12)\beta h\right)}
{\sinh\frac{\beta h}2}.
\label{resloop}
\end{equation}
For the $XXZ$ chain the corresponding object is obtained by substituting on
the rhs~of (\ref{resloop}) temperature $T\to\tilde T$, 
field $h\to\tilde h$ and spin $S\to1/2$.
The action of the QTM of the TL-model and that of the $XXZ$ chain in their
respective $N/2$-sectors are identical if
\begin{equation}
\frac{\sinh\left((S+\frac 12)\beta h\right)}{\sinh\frac{\beta h}2}=
\frac{\sinh\tilde\beta\tilde h}{\sinh\frac{\tilde\beta\tilde h}2}=
2\cosh\frac{\tilde\beta\tilde h}2.
\end{equation}
and the temperatures coincide $T=\tilde T$! Eventually we find the {\em
Temperley-Lieb equivalence for finite temperature and arbitrary field}
\begin{equation}
Z(T,h)=Z_{XXZ}(T,\tilde h) \quad\hbox{for} \quad
\frac{\sinh\left((S+\frac 12)\beta h\right)}{\sinh\frac{\beta h}2}=
2\cosh\frac{\beta \tilde h}2,
\label{partition}
\end{equation}
which is the generalization of (\ref{partition0}) to the case $h\not=0$.
The `identity' of the two partition functions only holds asymptotically, 
i.e.~$\lim_{L\to\infty}\left(Z/Z_{XXZ}\right)^{1/L}=1$. The identity holds
strictly for the free energies per site
\begin{equation}
f(T,h)=f_{XXZ}(T,\tilde h),
\label{freeenergy}
\end{equation}
with the relation of the magnetic fields and temperature given in
(\ref{partition}). (For the quantum $RSOS$ models, by use of the fusion
algebra, a similar relation was found in \cite{KlumperTherm92}. We
believe that (\ref{freeenergy}) is universally valid for all TL-models. However, a relation analogous to (\ref{partition}) for the
effective field $\tilde h$ is model dependent.)

\begin{figure}
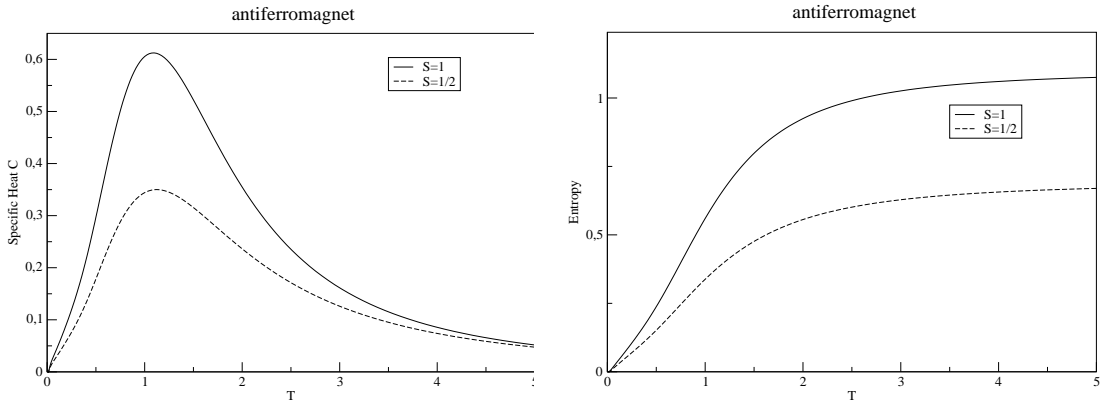

\begin{center}
\includegraphics[width=0.45\textwidth]{C_af2.eps}\quad 
\includegraphics[width=0.45\textwidth]{S_af2.eps}
\caption{\label{Fig1}Depiction of the temperature dependence of a) specific
  heat $c(T)$, and b) entropy $S(T)$ for the spin-1 (solid lines) and the
  spin-$1/2$ (dashed lines) quantum chains with antiferromagnetic exchange.}
\end{center}
\end{figure}

\begin{figure}
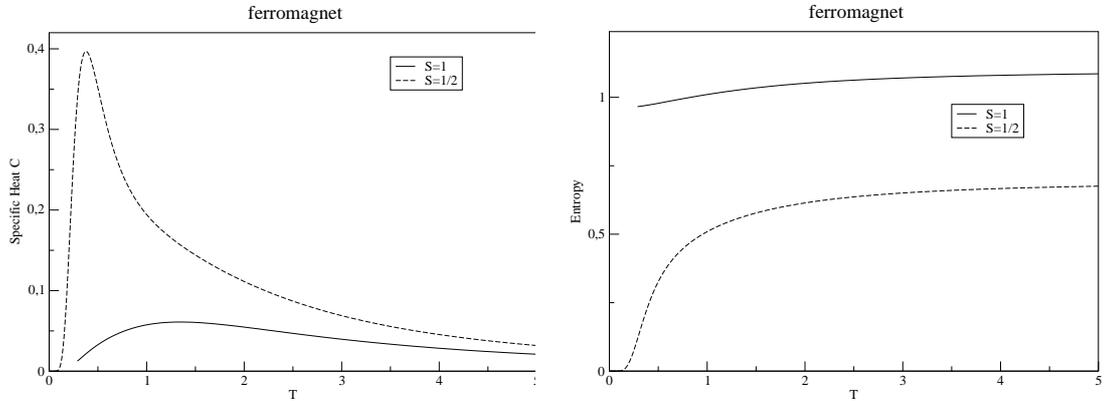

\begin{center}
\includegraphics[width=0.45\textwidth]{C_f2.eps}\quad 
\includegraphics[width=0.45\textwidth]{S_f2.eps}
\caption{\label{Fig2}Depiction of the temperature dependence of a) specific
  heat $c(T)$, and b) entropy $S(T)$ for the spin-1 (solid lines) and the
  spin-$1/2$ (dashed lines) quantum chains with ferromagnetic exchange. Note
  that data for the spin-1 chain were not calculated to very low temperatures,
  see the main text for details.}
\end{center}
\end{figure}

\begin{figure}
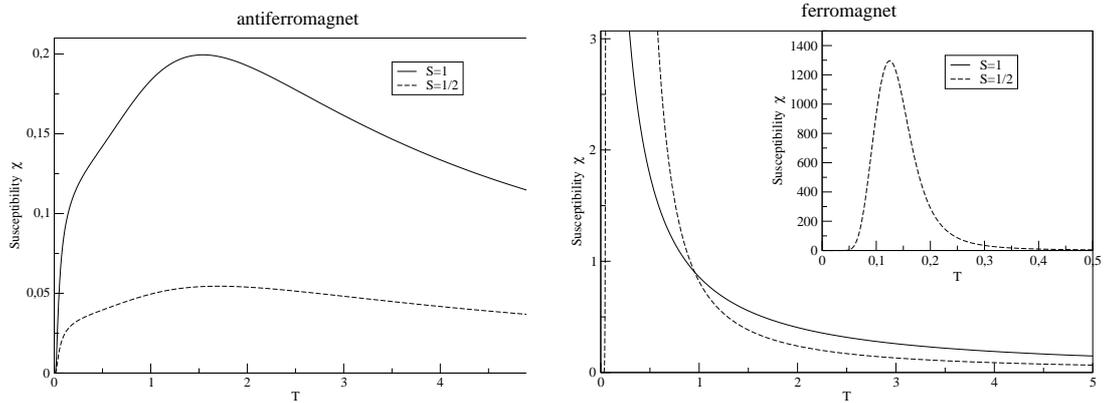

\begin{center}
\includegraphics[width=0.45\textwidth]{Chi_af2.eps}\quad 
\includegraphics[width=0.45\textwidth]{Chi_f2.eps}
\caption{\label{Fig3}Depiction of the temperature dependence of the magnetic
  susceptibility $\chi(T)$ in case of a) antiferromagnetic exchange and b)
  ferromagnetic exchange for the spin-1 (solid lines) and the spin-$1/2$
  (dashed lines) quantum chains. In the inset the full range of the
  susceptibility for the $XXZ$ chain is shown.}
\end{center}
\end{figure}

In figures~\ref{Fig1}-\ref{Fig3} we show zero-field results for specific heat
$c(T)$, entropy $S(T)$, and susceptibility $\chi(T)$ for the spin-1
biquadratic chain ($S=1$ TL-model) and the related $XXZ$ chain ($S=1/2$ with
$\Delta=3/2$) for antiferromagnetic and ferromagnetic signs of the exchange
coefficients. These results extend the already published results on the spin-1
biquadratic chain in \cite{KlumperTherm93}. The specific heat curves show a
finite temperature maximum and approach zero for $T\to 0$ and $T\to\infty$.
For the antiferromagnetic case, the specific heat data for the $S=1$ chain are
larger than those for the $S=1/2$ chain in agreement with the larger
integrated value of the reduced specific heat for the $S=1$ chain
\begin{equation}
\int_0^\infty \frac{c(T)}T dT=S(T=\infty)-S(T=0),
\label{IntegSpec}
\end{equation}
where $S(T)$ is the entropy. In the antiferromagnetic case, the entropy varies
monotonically from 0 to $\log 3$ ($\log 2$) for the spin-1 (spin-1/2) chain.
Note that the low temperature asymptotics show the usual thermodynamically
activated behaviour of gapped systems with an essential singularity. The gap
is actually rather small in the antiferromagnetic case $\Delta E_{af}=
0.173178...$ accompanied by a large correlation length $\xi= 21.0728505...$.

For the ferromagnetic case the numerical computations showed instabilities for
the spin-1 chain. We attribute these instabilities to purely numerical causes
and exclude physical reasons such as phase transitions. The data underlying the
illustrations are those which were obtained within reasonable computation
time.  The specific heat in the ferromagnetic case looks similar to the
antiferromagnetic case, however the order of the $S=1$ and the $S=1/2$ cases
is inverted. This seems to contradict (\ref{IntegSpec}) and the
high-temperature limits of the entropy $\log 3$ and $\log 2$. Note, however,
that in the case of the spin-1 biquadratic model, one of the rare special
cases with residual entropy is realized! Unlike the $S=1/2$ case and many
other systems, for the ferromagnetic spin-1 biquadratic model and others of
Sect.~\ref{TL-models} the ground state is exponentially degenerate! The
residual entropy is
\begin{equation}
S(T=0)=\log z=\log\left(\frac{d + \sqrt{d^2-4}}2\right),
\label{residentr}
\end{equation}
which gives $S(T=0)=\log\left(3+\sqrt 5\right)/2=0.9624...$ for dimension $d=3$ (and zero
for $d=2$).  This is also supported by the temperature dependence of the
entropy as shown in figure~\ref{Fig2}b). The value of the residual entropy for
$d=3$ was derived earlier \cite{EMuHa89}. Note that in the ferromagnetic
($S=1/2$) case, due to the larger excitation gap $\Delta E_f=2$, the
thermodynamically activated behaviour at low temperatures is better visible
than in the antiferromagnetic case.

In figure~\ref{Fig3} the susceptibility $\chi(T)$ data are presented. For
antiferromagnetic exchange coefficients, the $S=1$ and the $S=1/2$ cases
look similar. The susceptibilities show a finite temperature maximum and
approach zero for $T\to 0$ and $T\to\infty$. For the ferromagnetic case large
values are obtained by $\chi(T)$ at low temperatures. For $T\to 0$ however,
$\chi(T)$ drops to 0 again due to the finite excitation gap. This is observed
for the $S=1/2$ chain. Unfortunately, for $S=1$ the true low-$T$ behaviour is
not yet reached in the numerical treatment. On physical grounds, however, we
expect a drop of $\chi(T)$ to 0 for low $T$.

\section{Summary}
We have shown how to construct the irreducible invariant subspaces (sectors)
of Temperley-Lieb models in the case of open as well as periodic boundary
conditions. A central step in the construction of the sectors was the
identification of the one-dimensional representations of the (open as well as
periodically closed) Temperley-Lieb algebra for arbitrary chain length. 

The one-dimensional representations are also known as Bethe ansatz reference
states. In the periodically closed case, the reference states had to be
translationally invariant for being compatible with the boundary
conditions. The questions about the eigenvalues and multiplicities (!) of the
momentum operator in Hilbert spaces of tensor-product type and of reduced type
led to an interesting analysis with compact answer that we did not find in the
literature, but think should exist already.

The above findings lead to sobering insight with respect to alternative
approaches like the coordinate and the algebraic Bethe ansatz. The fact, that
most of the reference states of the Temperley-Lieb models with periodic
boundary conditions have non-zero momentum eigenvalues leads to the
equivalence with the $XXZ$ chain with twisted boundary conditions where the
twist is given by the momentum value. (Note that in an extreme case, also an
imaginary twist angle appears). Further, the higher spin-$S$ quantum chains
have exponentially degenerate ground-states. This should explain the failure
of attempts of direct Bethe ansatz calculations \cite{KS1994,KS1996} to construct
all eigenstates from just one standard reference state.

There are two types of applications of our results. We like to point out, that
the complete understanding of the spectrum of Temperley-Lieb systems with
periodic boundary conditions allows for a study of the conformal
dimensions. Here, we did not follow this line of thoughts and leave it for
future work. Most interesting with respect to applications in the theory of
the spin quantum Hall effect are vertex models with super-symmetry,
like $sl(n+1,n)$ and local $2n+1$ dimensional space corresponding to
$TL(\lambda=1)$.

As an application of the complete knowledge of multiplicities we
computed the thermodynamical properties of the quantum spin chains without
magnetic field by a direct mapping of the partition function to that of the
$XXZ$ chain. Such a treatment would have been possible already on the basis 
of results by \cite{Kulish03,KulishManojlovicNagy08}.
Interestingly, our alternative indirect approach to thermodynamics by taking
a detour via a classical two-dimensional model with twisted boundary
conditions allowed for a more transparent and more general treatment allowing
for arbitrary, non-vanishing (!) external fields. The result of these
investigations is a `Temperley-Lieb equivalence at finite temperature and
finite field'. The specific heat, entropy and susceptibility data of the
biquadratic model were explicitly calculated for arbitrary
temperature. Especially the low-temperature properties are very interesting.
In the ferromagnetic case the susceptibility data show large values at low
temperatures.

Our investigations are extensive, but not complete. We hope to report
elsewhere on a complete study of the low-temperature asymptotics of the
quantum spin chains and on the complete treatment of the non-semisimple cases
including the super-symmetric chains. Also, we expect a generalization of some
results to systems based on other algebras or formulated on lattices different
from a simple chain. In case of the Hecke algebra, the equivalent reference
models will be those solvable by nested Bethe ansatz.
\\

{\bf Acknowledgment:} B.A. acknowledges financial support by
the DFG research training group 1052 and by VolksWagen-Stiftung. The authors
are grateful to E. Müller-Hartmann, H.-A. Wischmann and C. Trippe for
scientific discussions.  The figures illustrating the thermodynamics of the
quantum chains were produced from data calculated by C. Trippe.


\begin{thebibliography}{1}


\bibitem{TemperleyLieb71} Temperley, H.N.V., Lieb, E.: Proc. R. Soc. Lond. A322, 251 
(1971)
\bibitem{BaxterBook82} Baxter R J 1982 Exactly Solved Models in Statistical Mechanics 
(London: Academic)


\bibitem{PMBook91}
P.~Martin: {\em {P}otts {M}odels and Related Problems in Statistical Mechanics}, 
World Scientific, (1991)

\bibitem{Jones83} Jones, V.R.F.: Invent. Math. 72, 1 (1983)


\bibitem{ABF84} Andrews, G.E, Baxter, R.J., Forrester, P.J. : J. Stat. Phys. 35, 
193 (1984)

\bibitem{OwczarekBaxter87} A.L. Owczarek and R.J. Baxter: {\it A class of 
interaction-round-a-face models and its equivalence with an ice-type model}, 
J. Stat. Phys. 49, 1093 (1987) 

\bibitem{Pasquier87} V. Pasquier, Nucl. Phys. B285 [FS19], 162 (1987); 
J. Phys. A 20, L1229 (1987); 5707 (1987)

\bibitem{KlumperBiq89} A. Kl\"umper: \it New results for $q$-state vertex models and the
pure biquadratic spin-1 Hamiltonian, \rm Europhys. Lett. \bf 9\rm, 815-820  (1989)

\bibitem{KulishReshetikhin81} P.~P.~Kulish and N. Reshetikhin: {\it Quantum linear problem for the sine-Gordon equation and higher representations}, Zap. Nauch. Semin. LOMI, 101, 101-110 (1981); Jour. Sov. Math. N4 (1983)

\bibitem{Kulish03} P.~P.~Kulish: {\it On spin systems related to the 
Temperley-Lieb algebra}, J. Phys. A: Math. Gen. 36 L489-L493 (2003) 

\bibitem{KulishManojlovicNagy08} P.~P.~Kulish, N. Manojlovic and Z. Nagy: {\it Quantum symmetry algebras of spin systems related to Temperley-Lieb R-matrices}, J. Math. Phys. 49 (2008)

\bibitem{Parkinson88} Parkinson J B: J . Phys. C: Solid State Phys. 20 L1029 (1987); J . Phys. C: Solid State Phys. 21 3793 (1988); J . Physique C8 {\bf 49}, 1413 (1988)

\bibitem{BarberBatchelor89} Barber M and Batchelor M T: Phys. Rev. B 40, 4621 (1989)

\bibitem{BazhanovReshetikhin89} Bazhanov, V. V. and Reshetikhin, N. Yu.: {\it Critical 
RSOS models and conformal field theory}, Int. J. Mod. Phys. A, 4 115-142 (1989)

\bibitem{PeriodicTL93}
P.~Martin and H.~Saleur:
{\it {O}n {A}n {A}lgebraic {A}pproach {T}o {H}igher {D}imensional
  {S}tatistical {M}echanics}, Commun. Math. Phys. 158, 155--190, (1993)

\bibitem{PeriodicTL94}
P.~Martin and H.~Saleur: {\it The blob algebra and the periodic Temperley-Lieb algebra}, Letters in 
Mathematical Physics 30, 189-206 (1994)

\bibitem{KlumperBiq90a} A. Kl\"umper: \it The spectra of $q$-state vertex models and 
related antiferromagnetic quantum spin chains. \rm J. Phys. A \bf 23\rm, 809-823 (1990)

\bibitem{KlumperBiq90b} A. Kl\"umper: \it Investigation of excitation spectra of exactly
solvable models using inverson relations. \rm Yang-Baxter Workshop/Conference,
Canberra, Int. J. Mod. Phys. B \bf 4\rm, 871 (1990)

\bibitem{AKS1992} Alcaraz, F. C., K\"oberle, R. and Lima-Santos, A.:
{\it All Exactly Solvable U(1)-Invariant Quantum Spin 1 Chains from Hecke Algebra},
Int. Journ. of Modern Physics A {\bf 7}, 7615 (1992)

\bibitem{KS1994} R.~K\"oberle and A.~Lima-Santos: {\it Exact solutions for the deformed biquadratic spin-1 chain} J.Phys. A27, 5409-5423 (1994)

\bibitem{KS1996} R.~K\"oberle and A.~Lima-Santos: {\it Exact solutions for {A}-{D} 
{T}emperley-{L}ieb models}, J. Phys. A: Math. Gen. 29, 519--531 (1996)

\bibitem{Levy91} D.~Levy: Phys.Rev.Lett. 67, 1971 (1991) 

\bibitem{PasquierSaleur89}
V.~Pasquier and H.~Saleur: {\it Common structures between finite systems and conformal field theories through Quantum Groups}, Nucl. Phys. B330, 523-556 (1990)

\bibitem{Goodman1989}
F.M. Goodman, P.~de~la Harpe, and V.F.R. Jones:
{\em Coxeter Graphs and Towers of Algebras}, Springer Verlag, (1989)

\bibitem{Westbury93}
B. W.~Westbury: {\it The representation theory of the Temperley-Lieb algebras}, 
Math. Z. 219, 539-565 (1995)

\bibitem{Nichols06}
A. Nichols: {\it The Temperley-Lieb algebra and its generalizations in the Potts and XXZ models}, J. Stat. Mech. (2006)

\bibitem{AlcarazBBBQ1987} F.C. Alcaraz, M.N. Barber, M.T. Batchelor, R.J. Baxter, 
and G.R.W. Quispel: {\it Surface exponents of the quantum {XXZ}, {A}shkin-{T}eller 
and {P}otts models}, J. Phys. A: Math. Gen. 20, 6397--6409 (1987)

\bibitem{SoerensenYoung90} Sørensen,~E.~S.; Young,~A.~P.: {\it Correlation
  length of the biquadratic spin-1 chain}, Phys. Rev. B 42, 754-759 (1990)

\bibitem{BatistaAligia2004} C. D. Batista and A. A. Aligia: {\it Exact Bond
  Ordered Ground State for the Transition between the Band and the Mott
  Insulator}, Phys. Rev. Lett. {\bf 92}, 246405 (2004)

\bibitem{AlcarazMalvezzi92} Alcaraz, F. C. and Malvezzi, A. L. - {\it On the 
Critical Behaviour of the Anisotropic Biquadratic Spin-1 Chain} - J. Phys. A: 
Math. Gen. {\bf 25}, 4535 (1992).

\bibitem{Albertini2000} G.~Albertini: {\it Is the purely biquadratic spin 1
  chain always massive?}  cond-mat/0012439 (December 2000)

\bibitem{KlumperTherm93} A.~Kl\"umper:
\it Thermodynamics of the anisotropic spin-1/2 Heisenberg chain and
related quantum chains \rm Z. Phys. B \bf 91\rm, 507-519 (1993)

\bibitem{KlumperTherm92} A.~Kl\"umper:
\it Free energy and correlation lengths of quantum chains related
to restricted solid-on-solid
lattice models \rm Ann. Phys. \bf 1\rm, 540-553 (1992)

\bibitem{EMuHa89} E.~M\"uller-Hartmann: unpublished results (1989)

\end{thebibliography}
\bibliographystyle{unsrt}

\end{document}